\documentclass[10pt]{article}
\usepackage{amssymb}
\usepackage{amsmath}
\usepackage{graphicx}
\usepackage{fancyhdr}
\usepackage{cite}

\begin{document}

\newcommand{\bin}[2]{\left(\begin{array}{c}\!#1\!\\\!#2\!\end{array}\right)}

\huge

\begin{center}
A quantitative study of some sources of uncertainty in opacity measurements
\end{center}

\vspace{0.5cm}

\large

\begin{center}
Jean-Christophe Pain\footnote{jean-christophe.pain@cea.fr} and Franck Gilleron
\end{center}

\normalsize

\begin{center}
CEA, DAM, DIF, F-91297 Arpajon, France
\end{center}

\vspace{0.5cm}

\begin{abstract}
Laboratory (laser and Z-pinch) opacity measurements of well-characterized plasmas provide data to assist inertial confinement fusion, astrophysics and atomic-physics research. In order to test the atomic-physics codes devoted to the calculation of radiative properties of hot plasmas, such experiments must fulfill a number of requirements. In this work, we discuss some sources of uncertainty in absorption-spectroscopy experiments, concerning areal mass, background emission, intensity of the backlighter and self-emission of the plasma. We also study the impact of spatial non-uniformities of the sample.  
\end{abstract}

\section{Introduction}\label{sec1}

In the introduction to the article ``Opacity calculations: past and future'' in 1964 \cite{MAYER64}, H. Mayer writes ``Initial steps for this symposium began a few billion years ago. As soon as the stars were formed, opacities became one of the basic subjects determining the structure of the physical world in which we live''. He is also the author of a famous report \cite{MAYER47} written in 1947 and presenting many theoretical methods for opacity calculation. 

Since Mayer's 1947 report, there have been few comprehensive reviews of methods for opacity calculations, although several reviews of limited scope have appeared, focusing on specific aspects \cite{HUEBNER14}. In early papers, the authors were mainly concerned with photo-ionization \cite{MARR67,BIBERMAN67,FANO68,STEWART67}. Many interesting summary articles about atomic opacities from the astrophysicists viewpoint were published \cite{COX65,CARSON68,CARSON71,CARSON72,HUEBNER86,ROGERS92}. Penner and Olfe discussed atomic and molecular opacities as applied to atmospheric reentry phenomena \cite{PENNER68} and the opacity of heated air was the subject of work by Armstrong \emph{et al.} \cite{ARMSTRONG61,ARMSTRONG67} and Avilova \emph{et al.} \cite{AVILOVA69,AVILOVA69b}. Proceedings of three opacity conferences were published by Mayer \cite{MAYER64}, Huebner et al. \cite{HUEBNER65} and Adelman and Wiese \cite{ADELMAN95}. Rickert \cite{RICKERT95} and Serduke \emph{et al.} \cite{SERDUKE00} summarized workshops in which opacities were compared. Efforts to calculate opacities and the underlying equations of state (EOS) have been renewed by the Los Alamos group, the Livermore group, and the opacity project at University College London and the University of Illinois. While the first two groups cover the entire range of atomic opacities, the last one concentrates on detailed EOS and opacities of light elements for stellar envelopes in the high temperature 3 10$^3$ $\leq T\leq$ 10$^7$ K and low density regions of the astrophysical plasma domain \cite{HUMMER91,SEATON55,BERRINGTON97a,BERRINGTON97b}. The low temperature limit avoids the presence of molecules and the high density limit is chosen so that the isolated atom or ion remains a reasonably good approximation.

Seventy-one years have passed, and despite the progress made, many researchers are still working on radiative opacity, due to its applications in the fields of inertial confinement fusion, defense and astrophysics. Several experiments were performed during the past three decades, but the recent experiment on the Z machine at Sandia National Laboratory (SNL), dedicated to the measurement of the transmission of iron in conditions close to the ones of the base of the convective zone of the Sun \cite{BAILEY15}, reveals that our computations may be wrong or at least incomplete.

The two main facilities used for opacity measurements are lasers and Z-pinches. High-power lasers and Z-pinches can be used to irradiate high-$Z$ targets with intense X-ray fluxes which volumetrically heat materials in local thermodynamic equilibrium (LTE) to substantial temperatures. These X-ray fluxes produce a state of high-energy density matter that can be studied by the technique of absorption spectroscopy. In such experiments, also known as pump-probe experiments, the X-ray source creating the plasma is expected to be Planckian.

The radiative quantities which are measured are the reference $I_{0,\nu}$ (unattenuated radiation intensity) and $I_{\nu}$, the radiation attenuated by the plasma. $\nu$ is the photon frequency. The transmission of the sample is given by 

\begin{equation}
\mathbb{T}_{\nu}=\frac{I_{\nu}}{I_{0,\nu}},
\end{equation}

\noindent and the knowledge of $\mathbb{T}_{\nu}$ and of the areal mass of the sample enables one to deduce the opacity (see section \ref{sec2}).

In laser experiments, the sample is heated by the X rays resulting from the conversion of the energy of laser beams focused inside a gold Hohlraum, and measurement of absorption coefficients in plasmas may be done for instance using the technique of point-projection spectroscopy, first introduced by Lewis and McGlinchey in 1985 \cite{LEWIS85}. The technique involves a small plasma produced by tightly focusing a laser on a massive or a fiber target to create a point-like X-ray source with a high continuum emission used to probe the heated sample. It was first used to probe expanding plasmas \cite{BRUNEAU91,CHENAIS89} and applied to probe a radiatively heated plasma for the first time in 1988 \cite{DAVIDSON88}. The point-projection spectroscopy technique can be used to infer the plasma conditions and/or its spectral absorption. 

Z-pinch experiments mentioned above proceed as follows \cite{BAILEY03,BAILEY07,BAILEY08,BAILEY09,NAGAYAMA19}. The process entails accelerating an annular tungsten Z-pinch plasma radially inward onto a cylindrical low density CH$_2$ foam, launching a radiating shock propagating toward the cylinder axis. Radiation trapped by the tungsten plasma forms a Hohlraum and a sample attached on the top diagnostic aperture is heated during a few nanoseconds when the shock is propagating inward and the radiation temperature rises. The radiation at the stagnation is used to probe the sample.

For a quantitative X-ray opacity experiment, great care must be taken in the preparation of the plasma, as the latter must be spatially uniform in both temperature and density. Masses and dimensions of the sample must be well-known. In order to compare with LTE opacity codes (see for instance \cite{COLGAN17,COLGAN18,PAIN15,PAIN17}), it is crucial to ensure that the plasma is actually in LTE. Quantitative information on the opacity can be obtained only if the following requirements are satisfied \cite{NAGAYAMA14,NAGAYAMA16,NAGAYAMA17}:

\begin{itemize}

\item (i) The instrumental spectral resolution has to be sufficiently high to resolve key line features and
measured accurately prior to the experiments.

\item (ii) Backlight radiation and tamper transmission have to be free of a wavelength-dependent structures.

\item (iii) Plasma self-emission has to be minimized.

\item (iv) The tamper-transmission difference has to be minimized.

\item (v) The sample conditions must be uniform, achieving near-local thermodynamic equilibrium.

\item (vi) The sample temperature, density, and drive radiation should be independently measured.

\item (vii) Measurements should be repeated with multiple sample thicknesses. 

\item (viii)  Both quantities $I_{0,\nu}$ and $I_{\nu}$ must be measured during the same experimental ``shot'' together with the plasma conditions.

\end{itemize}

The lack of simultaneous measurement of plasma conditions and absorption coefficient is a weakness of most absorption measurements. For example, some experiments rely on radiative-hydrodynamics simulations to infer the plasma temperature and density, while other provide measurements of temperature, density and absorption spectrum, but on different shots. However, even if many experimental teams devote lots of efforts to perform simultaneous measurements, the inferred quantities are always known with a limited accuracy. 

There are many sources of uncertainty: areal mass of the sample, background radiation, intensity of the backlighter, plasma temperature and density, etc. Strictly speaking, some of error quantification depends on platform (such as self-emission). Opacity-measurement uncertainty is challenging because it consists of three sources of errors that are complicated in different ways: 

\begin{itemize}

\item (i) transmission error, 

\item (ii) areal density error, and 

\item (iii) temperature and density errors. 

\end{itemize}

In addition, there are multiple sources for each category. When there is a lateral areal-density non-uniformity, effective areal density (and its error) become transmission dependent. Therefore, areal-density non-uniformity must be treated either as transmission error or as a special category. 

In the present work, we investigate some uncertainty sources in absorption spectroscopy measurements. In section \ref{sec2}, we show how the relative uncertainties on the transmission and on the areal mass $\rho L$ are related to the uncertainty on the opacity. The question we want to answer is: if we seek a particular value of the relative precision on opacity $\Delta\kappa_{\nu}/\kappa_{\nu}$, knowing the relative uncertainty on the areal mass $\Delta(\rho L)/(\rho L)$, which precision $\Delta\mathbb{T}_{\nu}$ on the transmission do we need? As an example, we impose the requirement $\Delta\kappa_{\nu}/\kappa_{\nu}$=10 \%. The uncertainties on the background emission and on the self-emission of the sample are discussed in section \ref{sec3}. We chose to discuss the latter areal-mass non-uniformities in a special category: different sources of uncertainty due to defects in the areal mass are examined in section \ref{sec4}: wedge shape, bulge (concave distortion), hollow (convex distortion), holes in the sample and oscillations. In  section \ref{sec5}, we address the issue of the temperature and density uncertainties. Section \ref{sec6} is the conclusion.

\section{Required precision on the measured transmission: error and uncertainty propagations}\label{sec2}

For a homogeneous and optically thin (non-emissive) material, the transmission is related to the opacity by the Beer-Lambert-Bouguer law \cite{BOUGUER1729,LAMBERT1760,BEER1852}:
 
\begin{equation}\label{beerlam}
\mathbb{T}_{\nu}=e^{-\rho L\kappa_{\nu}},
\end{equation}

\noindent where $\rho$ is the density and $L$ the thickness of the material, $\kappa_{\nu}$ its spectral opacity and $\mathbb{T}_{\nu}$ its transmission. $\rho L$ is the areal mass. Formula (\ref{beerlam}) is valid for $\rho L\kappa_{\nu}\lesssim 1$, \emph{i.e.} $\mathbb{T}_{\nu}\gtrsim 0.37$. 

First we have to specify what we mean by uncertainty. In particular, it is important to separate error and uncertainty. Error can be defined as the difference between measured value and the true value, $\Delta \mu=\mu_{\mathrm{meas}}-\mu_{\mathrm{true}}$, from a single measurement, which can go either positive or negative. On the contrary, uncertainty can be defined as interval (or width) of likelihood where a measured value could fall in. Usually, an uncertainty $\sigma$ given by an analysis or measurement represents a width of Gaussian probability distribution where the true value can be found. For example, if a measurement found $\mu_{\mathrm{meas}}\pm\sigma_{\mathrm{meas}}$, the true value $\mu_{\mathrm{true}}$ can be any value but its likelihood follows the Gaussian probability distribution defined as

\begin{equation}
\frac{1}{\sqrt{2\pi}\sigma_{\mathrm{meas}}}\exp\left[-\frac{\left(\mu-\mu_{\mathrm{meas}}\right)^2}{2\sigma_{\mathrm{meas}}^2}\right].
\end{equation} 

\noindent This is why it is usually considered that the true value exists within the measured $\mu\pm\sigma$ for 68 \% of the time. 

It is tempting to split the error propagation $\Delta\mathbb{T}_{\nu}\rightarrow\Delta\kappa_{\nu}$ in two separate steps: $\Delta\mathbb{T}_{\nu}\rightarrow\Delta\tau_{\nu}$ ($\tau_{\nu}=\rho L\kappa_{\nu}$ being the optical depth) and then $\Delta\tau_{\nu}\rightarrow\Delta\kappa_{\nu}$. The conversion from $\Delta\mathbb{T}_{\nu}$ to $\Delta\tau_{\nu}$ is complicated by nature due to their non-linear relation and $\mathbb{T}_{\nu}$ dependence. There are two main sources of $\Delta\mathbb{T}_{\nu}$: 

\begin{itemize}

\item (i) miscalibration between unattenuated intensity to attenuated intensity and 

\item (ii) background subtraction error. Plasma self-emission (\emph{i.e.}, sample and tamper) can be considered as a special case of background (see Sec. \ref{sec3}). 

\end{itemize}

The second phase is the conversion from $\Delta\tau_{\nu}$ to $\Delta\kappa_{\nu}$. This conversion is mathematically much less complicated than the $\Delta\mathbb{T}_{\nu}\rightarrow\Delta\tau_{\nu}$ one, and only areal-density errors $\Delta(\rho L)$ have to be quantified. For example, if $\rho L$ is perfectly known, the percent errors on $\tau_{\nu}$ and $\kappa_{\nu}$ are the same. If $\rho L$ is underestimated by 10 \%, opacity $\kappa_{\nu}$ ($=\tau_{\nu}/(\rho L)$) is (additionally) overestimated by 11 \% (\emph{i.e.}, 1/0.9 $\approx$ 1.11). This is well separable from $\Delta\mathbb{T}_{\nu}$ (or $\Delta\tau_{\nu}$). For example, if $\tau_{\nu}$ is overestimated by 10 \% due to transmission error (whatever the source is), 10 \%-underestimated $\rho L$ ends up in giving 1.1/0.9=1.22, which ends up in 22 \% overestimate in opacity. So, the impact of areal-density error can be separately computed from transmission (or optical-depth) error.

The quantity of interest \textit{in fine} is opacity $\kappa_{\nu}$, which is considered to be a function of areal mass $\rho L$ (measured by specific techniques such as Rutherford back-scattering for instance) and transmission $\mathbb{T}_{\nu}$. Actually, $\mathbb{T}_{\nu}$ is not measured directly; the quantities that are measured are transmitted intensity $I_{\nu}$, backlighter intensity $I_{0,\nu}$, electron density $n_e$, electron temperature $T$, etc.). Therefore, the proper way to study propagation error would be to consider $\kappa_{\nu}=f(n_e, L, T, I_{\nu}, I_{0,\nu}, \mathrm{etc.})$. In order to simplify the problem, we gather all these variables into two ones: areal mass $\rho L$ and transmission $\mathcal{T}_{\nu}=I_{\nu}/I_{0,\nu}$. Error propagation must absolutely be performed using $\kappa_{\nu}=f(\rho L, T_{\nu})$ and not $T_{\nu}=f(\rho L, \kappa_{\nu})$ (the latter procedure gives unrealistic uncertainties).

Carrying out the same measurement operation many times and calculating the standard deviation of the obtained values is one of the most common practices in measurement uncertainty estimation. Either the full measurement or only some parts of it can be repeated. In both cases useful information can be obtained. The obtained standard deviation is then the standard uncertainty estimate. If $q=f(x,y)$, the propagation of uncertainty reads \cite{TAYLOR97}:

\begin{equation}\label{quadi}
\sigma_q=\sqrt{\left[\frac{\partial f}{\partial x}\right]^2~\left(\sigma_x\right)^2+\left[\frac{\partial f}{\partial y}\right]^2\left(\sigma_y\right)^2},
\end{equation}

\noindent \emph{i.e.} if $q=x+y$:

\begin{equation}
\sigma_q=\sqrt{\left(\sigma_x\right)^2+\left(\sigma_y\right)^2}.
\end{equation}

\noindent In our case, $q=\kappa_{\nu}$, $x=\rho L$, $y=\mathbb{T}_{\nu}$, $f(x,y)=-\ln(y)/x$ and Eq. (\ref{quadi}) becomes

\begin{equation}\label{delkap}
\frac{\sigma_{\kappa_{\nu}}}{\kappa_{\nu}}=\sqrt{\left(\frac{\sigma_{\rho L}}{\rho L}\right)^2+\left(\frac{1}{\ln\mathbb{T}_{\nu}}\frac{\sigma_{\mathbb{T}_{\nu}}}{\mathbb{T}_{\nu}}\right)^2}.
\end{equation}

\noindent Uncertainty estimates obtained as standard deviations of repeated measurement results are called A-type uncertainty estimates. If uncertainty is estimated using some means other than statistical treatment of repeated measurement results then the obtained estimates are called B-type uncertainty estimates. The latter represent upper bounds on the variable of interest and have no statistical meaning. Therefore, if the quantity $z$ has the value $z_0$ with a relative uncertainty of $\Delta z/z$, it means that $z$ lies between $z_0-\Delta z$ and $z_0+\Delta z$, where $\Delta z$ is the absolute value of errors on any quantity $z$ and thus positive. The other means can be \emph{e.g.} certificates of reference materials, specifications or manuals of instruments, estimates based on long-term experience, \emph{etc}. The propagation of error bounds of a quantity $q$ depending on two independent variables $x$ and $y$ is

\begin{equation}\label{deltaq3}
\Delta_q=\left|\frac{\partial f}{\partial x}\right|\Delta_x+\left|\frac{\partial f}{\partial y}\right|\Delta_y.    
\end{equation} 

\noindent  If $q=x+y$, then $\Delta_q=\Delta_x+\Delta_y$. One has therefore

\begin{equation}\label{resuimp}
\frac{\Delta\kappa_{\nu}}{\kappa_{\nu}}=-\frac{1}{\ln\mathbb{T}_{\nu}}
\frac{\Delta\mathbb{T}_{\nu}}{\mathbb{T}_{\nu}}+\frac{\Delta(\rho L)}{\rho L}.
\end{equation}

\noindent Since Eq. (\ref{resuimp}) is the simple derivative, this is just a propagation of error while Eq. (\ref{delkap}) is the propagation of uncertainty \emph{stricto sensu}. To differentiate these two better, let us say we want to find $f$, which is a known function of $x$ and $y$, and we measure $x$ and $y$. Then, there are three cases: 

\begin{itemize}

\item If $x$ and $y$ are known perfectly, $f(x,y)$ is perfectly known.

\item If errors in $x$ and $y$ are known perfectly (\emph{i.e.}, $\Delta x$ and $\Delta y$, respectively), you can correct the errors in $f(x,y)$ perfectly using Eq. (\ref{resuimp}). 

\item If only uncertainties of $x$ and $y$ are known (\emph{i.e.}, $\sigma_x$ and $\sigma_y$ but not actual errors $\Delta x$ and $\Delta y$), one can only compute the likelihood interval (or the probability distribution) of true $f$ using $\sigma_f$ found with Eq. (\ref{delkap}). 

\end{itemize}

Sometimes, it is possible to know perfectly the uncertainty of one of the two variables $x$ or $y$, and to have an upper bound for the other. In such a case, neighther Eq. (\ref{delkap}) nor Eq. (\ref{resuimp}) are relevant for the propagation of errors or uncertainties. In this work, we therefore choose to always use Eq. (\ref{resuimp}) which is more restrictive than Eq. (\ref{delkap}). Using Eq. (\ref{resuimp}) one gets, in terms of error bars, the error on spectral transmission

\begin{equation}\label{deltatnu}
\Delta\mathbb{T}_{\nu}=-\mathbb{T}_{\nu}\ln(\mathbb{T}_{\nu})\left(\frac{\Delta\kappa_{\nu}}{\kappa_{\nu}}-\frac{\Delta(\rho L)}{\rho L}\right).
\end{equation}

\noindent In that case, errors represent upper bounds on the variables. The same result can be obtained using inequalities (\emph{i.e.} assuming that transmission lies between $\mathbb{T}_{\nu}-\Delta \mathbb{T}_{\nu}$ and $\mathbb{T}_{\nu}+\Delta \mathbb{T}_{\nu}$ and that areal mass lies between $\rho L-\Delta (\rho L)$ and $\rho L+\Delta (\rho L)$). The corresponding calculation is provided in Appendix A, since it might be helpful for didactic reasons (the connection between mathematical differentiation and positive variations $\Delta$ is hidden and not easy to understand). Eq. (\ref{deltatnu}) shows also that the error on the areal mass must not exceed the required precision on the opacity, which can be easily understood. 

One can also ``invert'' the formula (\ref{deltatnu}) in order to express the transmission that should be sought in order to ensure a given value of $\Delta\mathbb{T}_{\nu}$:

\begin{equation}\label{tnu}
\mathbb{T}_{\nu}=-\frac{\Delta\mathbb{T}_{\nu}}{\left(\frac{\Delta\kappa_{\nu}}{\kappa_{\nu}}-\frac{\Delta(\rho L)}{\rho L}\right)W\left(-\frac{\Delta\mathbb{T}_{\nu}}{\left[\frac{\Delta\kappa_{\nu}}{\kappa_{\nu}}-\frac{\Delta(\rho L)}{\rho L}\right]}\right)},
\end{equation}

\noindent where $W$ is Lambert's function ($w=W(z)$ is the solution of $we^w=z$, see Fig. \ref{fig_lambert}) \cite{LAMBERT1758,LAMBERT1772,CORLESS96,CAILLOL03,PAIN11,VEBERIC12,FUKUSHIMA13}. The Lambert function is named ``ProductLog'' in the Mathematica$^{ \textregistered}$ software. Table \ref{tab2} contains uncertainties mentioned in several publications about absorption-spectroscopy measurements over the past decades (see Refs. \cite{BAILEY03,FOSTER91,PERRY91,SPRINGER92,DASILVA92,EIDMANN94,PERRY96,MERDJI98,CHENAIS00,RENAUDIN06,LOISEL09,BLENSKI11a,BLENSKI11b}). For the enigmatic iron experiment on Z \cite{BAILEY15}, as well as for the more recent measurements on chromium, iron and nickel \cite{NAGAYAMA19}, the relative uncertainty on the areal mass $\Delta(\rho L)/(\rho L)$ was close to 4 \% (estimated from Rutherford back-scattering).  In a recent paper devoted to the ongoing National Ignition Facility (NIF) experiment on iron \cite{HEETER17}, Heeter \emph{et al.} estimate the relative uncertainty on the areal mass $\Delta(\rho L)/(\rho L)\approx 7 \%$. The definition of Lambert's function (for the principal branch $W_0$ in black in Fig. \ref{fig_lambert}) implies that:

\begin{equation}
W\left(-\frac{\Delta\mathbb{T}_{\nu}}{\left[\frac{\Delta\kappa_{\nu}}{\kappa_{\nu}}-\frac{\Delta(\rho L)}{\rho L}\right]}\right)\geq-1
\end{equation}

\noindent and

\begin{equation}
\frac{\Delta\mathbb{T}_{\nu}}{\left[\frac{\Delta\kappa_{\nu}}{\kappa_{\nu}}-\frac{\Delta(\rho L)}{\rho L}\right]}\leq\frac{1}{e},
\end{equation}

\noindent which means that the largest acceptable value of $\Delta\mathbb{T}_{\nu}$ corresponds to $\mathbb{T}_{\nu}=1/e\approx 0.37$. 
If we want a precision $\Delta\kappa_{\nu}/\kappa_{\nu}$ of 10 \% on the opacity, assuming a relative uncertainty of 7 \% on the areal mass, formula (\ref{deltatnu}) implies that, for a transmission $\mathbb{T}_{\nu}=0.4$, a precision of 0.011 is required on the transmission, which corresponds to  $\approx$ 2.75 \% (see table \ref{tab1} and Fig. \ref{fig_deltatnu}). Therefore, the value of 0.02 (5 \%) given by Heeter \emph{et al.} was a bit overestimated.

If $\Delta\rho L$ is perfectly known, in order to achieve $\Delta\kappa_{\nu}/\kappa_{\nu}=10 \%$, Eq. (\ref{deltatnu}) becomes 

\begin{equation}\label{lim1}
\frac{\Delta\mathbb{T}_{\nu}}{\mathbb{T}_{\nu}}=-0.1\times\ln(\mathbb{T}_{\nu}).
\end{equation}

\noindent Assuming $\Delta\rho L\ne 0$ ends up in tightening the $\Delta\mathbb{T}_{\nu}/\mathbb{T}_{\nu}$ measurement. For $\Delta(\rho L)/(\rho L)=7 \%$, Eq. (\ref{deltatnu}) becomes

\begin{equation}\label{lim2}
\frac{\Delta\mathbb{T}_{\nu}}{\mathbb{T}_{\nu}}=-0.03\times\ln(\mathbb{T}_{\nu}).
\end{equation}

\noindent Figure \ref{figt1t2} represents $\Delta\mathbb{T}_{\nu}/\mathbb{T}_{\nu}$ as a function of $\mathbb{T}_{\nu}$ in both cases (\ref{lim1}) and (\ref{lim2}).

We note that if we had used Eq. (\ref{delkap}) for uncertainty propagation, in order to achieve $\Delta\kappa_{\nu}/\kappa_{\nu}=10 \%$ with $\Delta(\rho L)/(\rho L)=7 \%$, we would have obtained, for $\mathbb{T}_{\nu}=0.4$, $\Delta\mathbb{T}_{\nu}=0.026$, which is larger than the estimate made using Eq. (\ref{resuimp}), \emph{i.e.} 0.011.

\subsection{Comparisons with Monte Carlo simulations}\label{subsec23}

Since Eq. (\ref{resuimp}) reflects a simple derivative, it is interesting to check the error propagation using Monte Carlo simulations. For given values of the transmission $\mathbb{T}_{\nu}$, as well as of the relative uncertainties $\Delta\mathbb{T}_{\nu}/\mathbb{T}_{\nu}$ and $\Delta(\rho L)/(\rho L)$, we have generated two sets of random numbers uniformly distributed, namely $X_i$ and $Y_i$ such as

\begin{equation}
1-\frac{\Delta(\rho L)}{\rho L}\leq X_i=\frac{\rho_iL_i}{\rho L}\leq 1+\frac{\Delta(\rho L)}{\rho L}
\end{equation}

\noindent and

\begin{equation}
1-\frac{\Delta\mathbb{T}_{\nu}}{\mathbb{T}_{\nu}}\leq Y_i=\frac{\mathbb{T}_i}{\mathbb{T}_{\nu}}\leq 1+\frac{\Delta\mathbb{T}_{\nu}}{\mathbb{T}_{\nu}}.
\end{equation}

\noindent Therefore, setting $\kappa_i=-\ln \mathbb{T}_i/\left(\rho_iL_i\right)$, one has

\begin{equation}
\frac{\Delta\kappa_{\nu}}{\kappa_{\nu}}=\max_{\mathrm{random}~i}\left(\frac{\Delta\kappa_{\nu}}{\kappa_{\nu}}\right)_i
\end{equation}

\noindent with

\begin{equation}
\left(\frac{\Delta\kappa_{\nu}}{\kappa_{\nu}}\right)_i=\frac{\left|\kappa_i-\kappa_{\nu}\right|}{\kappa_{\nu}}=\left|1-\frac{1}{X_i}\left(\frac{\ln Y_i}{\ln\mathbb{T}_0}+1\right)\right|,
\end{equation}

\noindent independent of the choice of $\rho L$ and $\kappa_{\nu}$. In the present case, we have chosen $\mathbb{T}_{\nu}$=0.4. Fig 2 shows the relative uncertainty on the opacity as a function of the transmission $\mathbb{T}_{\nu}$ for $\Delta (\rho L)/(\rho L)$=7\% and various values of $\Delta\mathbb{T}_{\nu}/\mathbb{T}_{\nu}$. As can be seen, in order to get a precision of 10 \% on the opacity for a transmission $\mathbb{T}_{\nu}$=0.4,  a precision of 2.75 \% is required on the transmission. This confirms the value obtained mentioned above.

\section{Background signal, backlighter characteristics and self-emission}\label{sec3}

\subsection{Uncertainty on the background signal}\label{subsec31}

Let us consider a wavelength-independent backlight signal $I_{0,\nu}\equiv I_0$ (the effect of wavelength-dependent structures was studied by Iglesias \cite{IGLESIAS06}). Assuming a background radiation of intensity $b_{\nu}$ in a restricted spectral region, we have

\begin{equation}\label{bgs}
\tilde{\mathbb{T}_{\nu}}=\frac{I_{\nu}-b_{\nu}}{I_0-b_{\nu}}.
\end{equation}

\noindent The quantity $\tilde{\mathbb{T}_{\nu}}$ represents the true transmission and $\mathbb{T}_{\nu}$ is the apparent transmission, \emph{i.e.} the transmission defined by Eq. (\ref{beerlam}) without substracting the background $\delta$. Expanding the latter expression up to second order yields

\begin{equation}
\tilde{\mathbb{T}_{\nu}}\approx\mathbb{T}_{\nu}\left(1-\frac{b_{\nu}}{\mathbb{T}_{\nu}I_0}\right)\left(1+\frac{b_{\nu}}{I_0}\right)=\mathbb{T}_{\nu}\left(1+\frac{b_{\nu}}{I_0}\left[1-\frac{1}{\mathbb{T}_{\nu}}\right]-\frac{b_{\nu}^2}{\mathbb{T}_{\nu}I_0^2}\right)+O(\delta^3).
\end{equation}

\noindent Therefore, at first order, we have

\begin{equation}
\tilde{\mathbb{T}_{\nu}}\approx\mathbb{T}_{\nu}\left(1+\epsilon\right)-\epsilon,
\end{equation}

\noindent where $\epsilon=b_{\nu}/I_0$. As can be seen in figures \ref{fig_back_1}, \ref{data6-background} and \ref{fig_back_2}, for a transmission of $\mathbb{T}_{\nu}\approx$ 0.4, if background level is as high as 10 \% of backlight intensity, transmission inferred without accounting for it (\emph{i.e.} using Eq. (\ref{beerlam}) instead of Eq. (\ref{bgs})) would significantly misinfer the sample transmission. This background error is bigger when expected transmission value is low (see Figs. \ref{fig_back_1} and \ref{fig_back_2}). This is the reason why it is required to repeat many experiments with varied sample thicknesses in order to measure strong lines at sufficiently high transmission. It is important to mention that Eq. (\ref{bgs}) can also be used to address the question of how the uncertainty $\Delta b_{\nu}$ on the background $b_{\nu}$ would impact the inferred opacity. Indeed, the modified transmission can be rewritten

\begin{equation}
\tilde{\mathbb{T}_{\nu}}=\frac{I_{\nu}-\left(b_{\nu}+\Delta b_{\nu}\right)}{I_0-\left(b_{\nu}+\Delta b_{\nu}\right)}=\frac{I_{\nu}'-\Delta b_{\nu}}{I_0'-\Delta b_{\nu}}
\end{equation}

\noindent with $I_{\nu}'=I_{\nu}-b_{\nu}$ and $I_0'=I_0-b_{\nu}$.

\subsection{Effect of self-emission}\label{subsec32}

The radiative-transfer equation for stationary, homogeneous and non-diffusive material reads

\begin{equation}\label{radtran}
\frac{dI_{\nu}}{dx}=-\rho\kappa_{\nu}I_{\nu}+j_{\nu},
\end{equation}

\noindent where $x$ represents the position along the line of sight of the spectrometer, $I_{\nu}$ is the intensity of the radiation field and $j_{\nu}$ the emissivity. For a plasma in LTE, using Kirchhoff's law $j_{\nu}=B_{\nu}\kappa_{\nu}$, where $B_{\nu}$ is the Planckian distribution

\begin{equation}
B_{\nu}=\frac{2h\nu^3}{c^2}\frac{1}{e^{\frac{h\nu}{k_BT}}-1},
\end{equation}

\noindent one gets the solution

\begin{equation}\label{blse}
\mathbb{T}_{\nu}=e^{-\rho L\kappa_{\nu}}+\frac{B_{\nu}}{I_0}\left(1-e^{-\rho L\kappa_{\nu}}\right).
\end{equation}

\noindent The transmission is therefore higher than the one predicted by Beer's law. However, Eq. (\ref{blse}) holds for a point-like source with a time-independent emission in one direction \cite{MIHALAS}, which is of course not representative at all of what really happens. The quantities $B_{\nu}$ and $I_0$ are usually given in erg/s/eV/cm$^2$/sr. However, the measured self-emission and backlight signals integrate $B_{\nu}$ and $I_0$, respectively, over their emitting area (observable from each point on the detector) and duration. Comparing these quantities without the integrations has limited applicability. In reality, the measurements also integrate over small energy range and solid angle as well, but for many platforms, the integrations over these quantities have similar impacts on backlight radiation and self-emission. A point-projection method \cite{PERRY17} has a greater chance of self-emission contamination. For example, if a 2 mm$^2$ sample foil is heated over 2.5 ns and backlit by a 200-$\mu$m backlight source over 300 ps, the ratio of sample-self-emission-to-backlight is close to 1/3 (in terms of expected photons per mm$^2$, see Figure 5 of Ref. \cite{ROSS16}). In the NIF experiment, there is no dedicated aperture \cite{ROSS16}, and thus, every point on the detector sees most of the emitting region. As a result, there is a huge difference in emitting area as well as in duration, which signifies the self-emission contamination relative to the backlight radiation. On the other hand, the Z-pinch experiment performed at SNL \cite{BAILEY15,NAGAYAMA19} uses a larger backlight area (800 $\mu$m), and the detector's view is limited by an aperture and a slit. As a result, the detector sees similar emitting surface areas for backlight radiation and sample self-emission (see the red rectangle of Fig. 16(a) of Ref. \cite{NAGAYAMA16}). In that way, the measurements are less subject to self-emission issues than the laser experiments. The durations of self-emission and backlight radiation are similar too, because the same source works as heating and backlight radiation. This might be the reason why NIF experiment suffers from self-emission and background, on the contrary to the SNL experiment. However, one has to be cautious; the background can also originate from other sources (such as Hohlraum itself, crystal second- (or higher-) order reflection, some sort of fluorescence or hard X rays, etc.). In order to quantify the impact of self-emission on the opacity measurements, one should consider the differences in the emission areas as well as the integration over their time histories (see Sec. IV-C of Ref. \cite{NAGAYAMA17}).
 
It is worth mentioning that the point-projection method evoked in the introduction is not the best method at high temperature due to this reason. 

\section{Modeling spatial non-uniformities of the sample}\label{sec4}

Figure \ref{defects} represents microscopy views of several copper samples (before the experiment) used during a recent (2017) experimental campaign in ``Laboratoire pour l'Utilisation des Lasers Intenses'' (LULI) in France \cite{POIRIER18}. As can be seen, the surface of the sample is far from being perfect. Spatial non-uniformities of targets have been widely investigated in the past (see for instance the non-exhaustive list of references \cite{YAAKOBI00,DEFRIEND11,ZHANG12}). 
There are various possible modulations. In reality, some modulation that exists in the target fabrication is relaxed during the experiment, while some other modulation might be produced by hydrodynamics. In the present work, we focus on five spatial deformations of the target in two dimensions: wedge, bulge (convex deformation), hollow (concave deformation), holes and oscillations (modulations of the surface), assuming that they are still present at the time of probe. Our goal is to find a simple analytical modeling of such non-uniformities in order to get a realistic idea of their respective impact.

The areal mass, written $\rho L$, can have defects in both directions $x$ and $y$ (we do not, for simplicity, separate the variations of $\rho$ and $L$ with respect to $x$ and $y$ but $\rho L$ is taken as a global quantity) and its average reads

\begin{equation}\label{23}
\langle\rho L(x,y)\rangle=\frac{1}{ab}\int_0^a\int_0^b\rho L(x,y)dxdy,
\end{equation}

\noindent where $a$ and $b$ are the dimensions of the sample in directions $x$ and $y$ respectively.

Sometimes, the interpretation of an experimental spectrum reveals that the main structures have the right energy and relative intensities which seem consistent with the experiment, but the general level of transmission is not satisfactory. Even if the areal mass of the sample is guaranteed by the manufacturer with a good accuracy, it might happen that some variations of the areal mass occur, during the experiment. There might be a difference between the areal density used in the analysis and the true areal density. In the following, we assume that area-averaged areal density $\langle\rho L(x,y)\rangle$ during the experiment is known and defined as Eq. (\ref{23}) and try to quantify how different types of lateral variations would affect the opacity inferred with $\langle\rho L(x,y)\rangle$.

In the present work, $\rho L$ is considered as a function of lateral (or transverse) position $x$ only (we still do not separate the dependence of $\rho$ and $L$ with respect to $x$ and take $\rho L$ as a global quantity) and in both cases, we preserve the average areal mass:

\begin{equation}\label{pres}
\langle\rho L(x)\rangle=\frac{1}{a}\int_0^a\rho L(x)dx=\rho_0L_0,
\end{equation}

\noindent $a$ being the transverse dimension of the target. For a perfect sample (no defects), we have

\begin{equation}
\tilde{\mathbb{T}_{\nu}}=\mathbb{T}_{\nu}=e^{-\kappa_{\nu}\rho_0L_0}
\end{equation}

\noindent and for a corrugated sample, we have

\begin{equation}\label{corrsam}
\tilde{\mathbb{T}_{\nu}}=\frac{1}{a}\int_0^ae^{-\kappa_{\nu}\rho L(x)}dx.
\end{equation}

\noindent Equation (\ref{corrsam}) is correct only when backlight radiation is uniformly filling the observed sample area. For NIF, it should not be a problem since backlight is only bright over 100 $\mu$m (although it has a self-emission issue). The concern is modulation over 100 $\times$ 100 $\mu$m$^2$ backlit region. For SNL, it is a bigger concern since backlighter is bigger (approximately 800 $\mu$m). However, it should not be a serious problem because the measurement resolves in one direction and takes lineout only over brightest 300 $\mu$m. Since it does not resolve in other direction, the modulation concern for SNL experiment is over 800 (backlight width) $\times$ 300 (lineout width) $\mu$m$^2$ region. The convexity of the exponential function implies

\begin{equation}
\langle e^{-\kappa_{\nu}\rho L(x)}\rangle\geq e^{-\kappa_{\nu}\langle\rho L(x)\rangle}
\end{equation}

\noindent and therefore, due to Eq. (\ref{pres}), we have

\begin{equation}
\tilde{\mathbb{T}_{\nu}}\geq\mathbb{T}_{\nu}.
\end{equation}

\noindent Assuming constant areal mass, the presence of spatial non-uniformities or distortions of the sample tends to make the foil more transparent. This is correct but may mislead the reader to think effective areal density is always lower. This is not true if the sample is tilted somehow by an angle $\theta$, maybe due to misalignment or some weird hydrodynamics. If tilt happens, the apparent areal density along the line of sight is elongated by $1/\cos(\theta)$. We have considered a few distortions:
 
\begin{itemize}
 
\item Wedge: in order to model a wedge-shape distortion of the sample (see Fig. \ref{wedge}), we use the linear form
 
\begin{equation}
\rho L(x)=\rho_1L_1\left(1-\epsilon\frac{x}{a}\right),
\end{equation}

\noindent where $\epsilon$ controls the slope of the surface of the target. 

\item Bulge: a concave distortion of the sample (see Fig. \ref{bulge}) is modeled here as

\begin{equation}
\rho L(x)=\rho_1L_1\left[-\frac{4\epsilon}{a^2}x(x-a)+1\right].
\end{equation}

\item Hollow: a convex distortion (see Fig. \ref{hollow}) is modeled here as

\begin{equation}
\rho L(x)=\rho_1L_1\left[\frac{4\epsilon}{a^2}x(x-a)+1\right].
\end{equation}

\item Impact of holes: we choose to model the presence of holes (see Fig. \ref{holes}) in the sample by the replacement

\begin{equation}
\rho_0L_0 \rightarrow\rho_1L_1(1-\epsilon),
\end{equation}

\noindent where $0<\epsilon<1$ quantifies the amount of holes (the areal mass of the sample must therefore be increased from $\rho_0L_0$ to $\rho_1L_1$). 

\end{itemize}

The different expressions of the modified transmission are summarized in table \ref{tab3} and the derivations are provided in Appendix B. The effect of the holes is stronger than the effect of the thickness modulations (25 \% of holes yield the same result as 100 \% of modulations). Nevertheless, to have a visible effect, one needs around 75-100 \% of spatial modulations of the areal mass; it seems unrealistic to have such a perturbed hydrodynamic evolution and / or such a bad conception of the targets. 

The impact of the different defects, as $\epsilon$ varies, is illustrated respectively in Figs. \ref{fig_wedge_1}, \ref{data1-wedge} and \ref{fig_wedge_2} (wedge), Figs. \ref{fig_bulge_1}, \ref{data2-bulge} and \ref{fig_bulge_2} (bulge), Figs. \ref{fig_hollow_1}, \ref{data3-hollow} and \ref{fig_hollow_2} (hollow), Figs. \ref{fig_holes_1}, \ref{data4-holes} and \ref{fig_holes_2} (holes) and Figs. \ref{fig_oscillations_1}, \ref{data5-modulations} and \ref{fig_oscillations_2} (modulations). Values of $\epsilon$ required to obtain an uncertainty of 7 \% on the areal mass (as in the NIF experiment \cite{HEETER17,PERRY17,HEETER18}) are displayed in table \ref{tab4}.

It is difficult for us to clarify what causes each type of non-uniformity (e.g., target fabrication, non 1-D expansion during experiment, instabilities during experiments). Of course, if the temperature is sufficiently high (which is the case in the laser or Z-pinch experiments mentioned above), the sample becomes a plasma, and the inhomogeneities will not be the same as the ones before the experiment (in the solid phase). For instance, one can imagine that the holes will be filled very quickly. In fact, the above considerations imply that we consider the defects of the sample at the instant of the probe. 

\section{Temperature and density errors}\label{sec5}

\subsection{A complex issue}\label{subsec51}

The temperature and density errors do not contribute to the opacity measurement itself. They are important only when comparing with models. Even if temperature and density are off by 50 \%, it would be fine if the calculations were identical at the misinferred conditions. In fact, temperature and density uncertainties are very different from $\Delta\mathbb{T}$ and $\Delta(\rho L)$ ones, and cannot be discussed in a general way. The criteria must depend on the level of model-data discrepancies. The relevant question for this uncertainty is: can the observed discrepancy between measurement and modeling be explained by temperature and density uncertainties?

\subsection{Upper bound on the uncertainty on temperature and density due to the cross-section calculation}\label{subsec52}

Uncertainties in opacity calculations stem from the fundamental atomic cross-sections, plasma effects caused by perturbing ions, computational limitations, etc. Measurements of fundamental cross-sections are usually carried out on neutral atoms, rather than on charged ions, due to the difficulty in preparing a sample in a specific ion stage and because of the myriad possibilities of excited levels. The problem is that cross-sections of neutral atoms are more difficult to calculate accurately because of the many-body electron-electron interaction. Thus, comparison of calculations with measured cross-sections for neutral species should provide an upper bound on uncertainties. Huebner and Barfield \cite{HUEBNER14} estimate that:

\begin{itemize}

\item (i) When scattering dominates (high temperature, low density), the uncertainty in the opacity is 5 \%. 

\item (ii) As the density increases, free-free processes become more important, the uncertainty is less than 10 \%. 

\item (iii) As the temperature decreases and bound-free processes become important, the uncertainty increases to 15-20 \% and as the temperature decreases still further (photo-excitation can contribute), the uncertainty increases to 30 \%. 

\end{itemize}

The calculated opacity error purely due to electron temperature  and density errors is not just cross-section error, but it involves cross-sections (oscillator strengths) combined with  error on quantum-state populations as well as on line broadening (in the case of photo-excitation) and edge broadening (in the case of photo-ionization). Since we are mostly dealing with LTE opacities in the present work, the population factor reduces to a Boltzmann factor, and depends on the energies of the configurations (and therefore in particular on the way electron-electron interactions are taken into account). Line shapes depend on many factors such as the atomic-physics basis used in the computation, the microfield distribution, etc. In the present subsection, we try to find upper bounds on temperature and density uncertainties required to ensure a given relative uncertainty on opacity; for that purpose we simplify the problem as much as we can, and restrict ourselves to one process (the simplest one to model approximately), namely inverse Bremsstrahlung. Kramers' formula for inverse Bremsstrahlung reads:

\begin{equation}
\kappa_{\nu}\approx\left(\frac{\mathcal{N}_A}{A}\right)^2\frac{32\pi^3}{3\sqrt{3}}\left(\frac{e^2}{4\pi\epsilon_0}\right)^3\frac{\hbar^2}{mc\sqrt{2\pi mk_BT}}\frac{Z^{*3}\rho}{\left(h\nu\right)^3},
\end{equation}

\noindent which can be put in the form $\kappa_{\nu}\approx CT^{-1/2}\rho/\nu^3$ yielding

\begin{equation}\label{inckra}
\frac{\Delta\kappa_{\nu}}{\kappa_{\nu}}\approx\frac{1}{2}\frac{\Delta T}{T}+\frac{\Delta\rho}{\rho}.
\end{equation}

\noindent Of course, this is not realistic; in the experiments, our main goal is to study plasmas for which the photo-excitation is important. In general, it is difficult to find a simple scaling with density and temperature. If we require $\Delta\kappa_{\nu}/\kappa_{\nu}\approx 0.1$, relation (\ref{inckra}) implies that

\begin{equation}
\frac{\Delta T}{T}<2\times 0.1=20~\%.
\end{equation}

\noindent and

\begin{equation}
\frac{\Delta\rho}{\rho}<10~\%.
\end{equation}

\noindent Thus, if observed discrepancy is 10 \%, $T$ and $n_e$ need to be known better than 20 \% and 10 \%. This is a crude approximation, but since it is reasonable to assume that $d\kappa_{\nu}/dT$ and $d\kappa_{\nu}/dn_e$ are similar between models than $\kappa_{\nu}$ itself, such estimates should be relevant.

\subsection{A remark on temperature inhomogeneities}\label{subsec53} 

Recently, Busquet \cite{BUSQUET17} paid attention to the fact that under some circumstances, the transmission of a L- or M-shell weakly inhomogeneous plasma is identical to the transmission of a one-temperature plasma. This is clearly demonstrated in the case of an opacity varying linearly with the temperature. Indeed, if

\begin{equation}
\mathbb{T}_{\nu}=\exp\left\{-\int_{y_1}^{y_2}\kappa_{\nu}\left[T(y),\rho(y)\right]\rho(y)dy\right\},
\end{equation}

\noindent where $T(y)$ and $\rho(y)$ are temperature and density of the sample at depth $y$, $y_1$ and $y_2$ being the limits of the sample (see Fig. \ref{gradients}). Defining the average temperature $\bar{T}$ as

\begin{equation}
\bar{T}=\int_{y_1}^{y_2}T(y)\rho(y) dy/\int_{y_1}^{y_2}\rho(y) dy,
\end{equation}

\noindent Busquet assumed a uniform density $\rho$ (but this is not necessary), and a linear dependence of $\kappa_{\nu}$ with respect to temperature:

\begin{equation}
\kappa_{\nu}(T,\rho)=\kappa_{\nu}(\bar{T},\rho)+(T-\bar{T})\times d,
\end{equation}

\noindent where 

\begin{equation}
d=\left.\frac{d\kappa_{\nu}}{dT}\right|_{T=\bar{T}}
\end{equation}

\noindent and therefore

\begin{equation}
\mathbb{T}_{\nu}=\exp\left[-\int_{y_1}^{y_2}\kappa_{\nu}(\bar{T})\rho(y)dy\right],
\end{equation}

\noindent which means that $\mathbb{T}_{\nu}$ is identical to the transmission of a sample at the average temperature. This implies that, under particular circumstances, one can find a temperature  for which the experiment can be interpreted, although the plasma is subject to gradients... This enforces the need for independent diagnostics of the plasma conditions (K-shell spectroscopy of a light element, Thomson scattering, shadowgraphy, etc.).

\section{Conclusion}\label{sec6}

It is important to study the effect of relative uncertainties in photo-absorption measurements (areal mass, backlighter, background radiation, self-emission, etc.). In the present work, we discussed, assuming the knowledge of the uncertainty on the areal mass, the required uncertainty on the transmission measurement in order to infer the opacity with a given accuracy. The issue of the uncertainty on the backlighter emission was also briefly investigated. We quantified the impact of several spatial non-uniformities of the areal mass on the transmission, considering fives cases: wedge, bulge, hollow, holes and modulations. The corresponding formulas can provide an insight on the effect of the amount of corrugations, depending in each case on a single parameter $\epsilon$ (\emph{e.g.} depth of the hollow relative to the nominal thickness, density of holes, amplitude of modulations, etc.). Non-uniformities of the sample can be detected by analysis techniques (Rutherford back-scattering, scanning electron microscopy, etc.). Such defects always overestimate the transmission, \emph{i.e.} make the plasma more transparent, due to convexity of the exponential function. They are more important when the spectral transmission is low. Holes, modulations (oscillations) and hollows (convex distortions) are expected to have the strongest impact.

\section{Appendix A: Expression of opacity error in terms of areal-mass and transmission relative uncertainties}

Let us denote $\kappa$ the opacity, $\mathbb{T}$ the transmission and $\mathcal{A}$ the areal mass. One has:

\begin{equation}
\mathbb{T}_{\nu}-\Delta \mathbb{T}_{\nu} \leq \mathbb{T} \leq \mathbb{T}_{\nu}+\Delta \mathbb{T}_{\nu} 
\end{equation}

\begin{equation}
\mathbb{\kappa}_{\nu}-\Delta \kappa_{\nu} \leq \kappa \leq \kappa_{\nu}+\Delta \kappa_{\nu} 
\end{equation}

\begin{equation}
\rho L-\Delta (\rho L) \leq \mathcal{A}\leq \rho L+\Delta (\rho L) 
\end{equation}

\noindent One has therefore

\begin{equation}
-\ln\left[\mathbb{T}_{\nu}-\Delta \mathbb{T}_{\nu}\right] \geq -\ln \mathbb{T}\geq -\ln\left[\mathbb{T}_{\nu}+\Delta \mathbb{T}_{\nu}\right] 
\end{equation}

\noindent and

\begin{equation}
\frac{1}{\rho L-\Delta (\rho L)} \geq \mathcal{A} \geq \frac{1}{\rho L+\Delta (\rho L)}, 
\end{equation}

\noindent which implies

\begin{equation}
\frac{-\ln\left[\mathbb{T}_{\nu}-\Delta \mathbb{T}_{\nu}\right]}{\rho L-\Delta (\rho L)} \geq \kappa \geq \frac{-\ln\left[\mathbb{T}_{\nu}+\Delta \mathbb{T}_{\nu}\right]}{\rho L+\Delta (\rho L)}. 
\end{equation}

\noindent Setting

\begin{equation}\label{kappamin}
\kappa_{\mathrm{min}}=\frac{-\ln\left[\mathbb{T}_{\nu}+\Delta \mathbb{T}_{\nu}\right]}{\rho L+\Delta (\rho L)}=\kappa_{\nu}-\Delta\kappa_{\nu}
\end{equation}

\noindent and 

\begin{equation}
\kappa_{\mathrm{max}}=\frac{-\ln\left[\mathbb{T}_{\nu}-\Delta \mathbb{T}_{\nu}\right]}{\rho L-\Delta (\rho L)}=\kappa_{\nu}+\Delta\kappa_{\nu}
\end{equation}

\noindent where $\kappa_{\nu}=-\ln\mathbb{T}_{\nu}/(\rho L)$, we have

\begin{equation}
\kappa_{\mathrm{max}}\approx-\frac{\ln \mathbb{T}_{\nu}}{\rho L}\left(1+\frac{\Delta (\rho L)}{\rho L}\right)-\frac{1}{\rho L}\ln\left[1-\frac{\Delta \mathbb{T}_{\nu}}{\mathbb{T}_{\nu}}\right]\left[1+\frac{\Delta(\rho L)}{\rho L}\right]
\end{equation}

\noindent and taking only the first-order terms in $\Delta \mathbb{T}_{\nu}$ and $\Delta(\rho L)$ and neglecting the second-order terms in $\Delta \mathbb{T}_{\nu}\Delta(\rho L)$:

\begin{equation}
\kappa_{\mathrm{max}}=\kappa_{\nu}+\Delta\kappa_{\nu}\approx\kappa_{\nu}+\kappa_{\nu}\frac{\Delta (\rho L)}{\rho L}+\frac{1}{\rho L}\frac{\Delta \mathbb{T}_{\nu}}{\mathbb{T}_{\nu}}
\end{equation}

\noindent or

\begin{equation}
\frac{\Delta\kappa_{\nu}}{\kappa_{\nu}}\approx\frac{\Delta(\rho L)}{\rho L}-\frac{1}{\ln \mathbb{T}_{\nu}}\frac{\Delta \mathbb{T}_{\nu}}{\mathbb{T}_{\nu}},
\end{equation}

\noindent which is exactly Eq. (\ref{resuimp}). The same result can of course be obtained using $\kappa_{\mathrm{min}}$ (see Eq. (\ref{kappamin})).

\section{Appendix A: Impact of areal-mass distortions on the transmission - analytical formulas}

In this appendix, we provide the main steps of the derivations of expressions givven in table \ref{tab3}.

\subsection{Wedge}\label{subseca1}

In order to model a wedge-shape distortion of the sample (see Fig. \ref{wedge}), we use the linear form
 
\begin{equation}
\rho L(x)=\rho_1L_1\left(1-\epsilon\frac{x}{a}\right),
\end{equation}

\noindent where $\epsilon$ controls the slope of the surface of the target. The preservation of the areal mass reads

\begin{equation}
\rho_1L_1\frac{1}{a}\int_0^a\left(1-\epsilon\frac{x}{a}\right)dx=\rho_0L_0,
\end{equation}

\noindent which yields

\begin{equation}
\rho_1L_1=\frac{\rho_0L_0}{1-\epsilon/2}
\end{equation}

\noindent and we get

\begin{equation}
\tilde{\mathbb{T}_{\nu}}=\frac{\left(1-\epsilon/2\right)}{\epsilon\ln\left(\mathbb{T}_{\nu}\right)}\left[\mathbb{T}_{\nu}\right]^{\frac{1}{1-\epsilon/2}}\left[1-\mathbb{T}_{\nu}^{-\frac{\epsilon}{1-\epsilon/2}}\right].
\end{equation}

\subsection{Bulge (convex distortion)}\label{subseca2}

A bulge-shape (concave) distortion of the sample (see Fig. \ref{bulge}) is modeled here as

\begin{equation}
\rho L(x)=\rho_1L_1\left[-\frac{4\epsilon}{a^2}x(x-a)+1\right].
\end{equation}

\noindent The quantity $\rho_1L_1$, given by the preservation of the areal mass, is

\begin{equation}
\rho_1L_1=\frac{\rho_0L_0}{1+2\epsilon/3}.
\end{equation}

\noindent We have therefore

\begin{equation}
\tilde{\mathbb{T}}_{\nu}=\sqrt{\frac{1+2\epsilon/3}{-\epsilon\ln\mathbb{T}_{\nu}}}~\left[\mathbb{T}_{\nu}\right]^{\frac{1}{1+2\epsilon/3}}~D\left(\sqrt{\frac{-\epsilon\ln\mathbb{T}_{\nu}}{1+2\epsilon/3}}\right),
\end{equation}

\noindent where $D(x)$ represents Dawson's function

\begin{equation}
D(x)=e^{-x^2}\int_0^x e^{t^2}dt.
\end{equation}

\subsection{Hollow (convex distortion)}\label{subseca3}

In a similar way, a hollow (convex distortion, see Fig. \ref{hollow}) is modeled here as

\begin{equation}
\rho L(x)=\rho_1L_1\left[\frac{4\epsilon}{a^2}x(x-a)+1\right].
\end{equation}

\noindent The thickness $L_1$, given by the preservation of the areal mass, is

\begin{equation}
\rho_1L_1=\frac{\rho_0L_0}{1-2\epsilon/3}.
\end{equation}

\noindent In the case of a hollow, we have

\begin{equation}
\tilde{\mathbb{T}}_{\nu}=\frac{\sqrt{\pi}}{2}\left[\mathbb{T}_{\nu}\right]^{\frac{1-\epsilon}{1-2\epsilon/3}}~\sqrt{\frac{1-2\epsilon/3}{-\epsilon\ln\mathbb{T}_{\nu}}}~\mathrm{Erf}\left[\sqrt{\frac{-\epsilon\ln\mathbb{T}_{\nu}}{1-2\epsilon/3}}\right],
\end{equation}
 
\noindent where $\mathrm{Erf}$ is the usual error function

\begin{equation}
\mathrm{Erf}(x)=\frac{2}{\sqrt{\pi}}\int_0^xe^{-t^2}dt.
\end{equation}

\noindent Nevertheless, to have a visible effect, at least 25 \% of holes are required in the target (this is very important, but since the considered thicknesses are of the order of a few hundreds of Angstr\"oms, it might be realistic).

\subsection{Impact of holes}\label{subseca4}

We choose to model the presence of holes (see Fig. \ref{holes}) in the sample by the replacement

\begin{equation}
\rho_0L_0 \rightarrow\rho_1L_1(1-\epsilon),
\end{equation}

\noindent where $0<\epsilon<1$ quantifies the amount of holes (the areal mass of the sample must therefore be increased from $\rho_0L_0$ to $\rho_1L_1$). The preservation of areal mass implies

\begin{equation}
\rho_1L_1=\frac{\rho_0L_0}{1-\epsilon}
\end{equation}

\noindent and the transmission becomes

\begin{equation}
\tilde{\mathbb{T}_{\nu}}=(1-\epsilon)\left[\mathbb{T}_{\nu}\right]^{\frac{1}{1-\epsilon}}+\epsilon.
\end{equation}

\subsection{Modulations}\label{subseca5}

We consider the following modulations of the sample (see Fig. \ref{oscillations}):

\begin{equation}
\rho L(x)=\rho_0L_0\left[1+\epsilon\cos(x)\right]
\end{equation}

\noindent with $0\leq\epsilon\leq 1$. We have, taking $a=2\pi N$ with $N\in \mathbb{N}$:

\begin{equation}
\langle\rho L(x)\rangle=\rho_0L_0\frac{1}{a}\int_0^a\left[1+\epsilon\cos(x)\right]dx=\rho_0L_0\left(1+\frac{\epsilon}{a}\sin(a)\right)=\rho_0L_0.
\end{equation}

\noindent We get

\begin{equation}
\tilde{\mathbb{T}_{\nu}}=\mathbb{T}_{\nu}\times\frac{1}{2\pi N}\int_0^{2\pi N}\left(\mathbb{T}_{\nu}\right)^{\epsilon\cos(x)}dx=\mathbb{T}_{\nu}\times\frac{1}{2\pi}\int_0^{2\pi}e^{\epsilon\cos(x)\ln\left(\mathbb{T}_{\nu}\right)}dx
\end{equation}

\noindent and finally

\begin{equation}
\tilde{\mathbb{T}_{\nu}}=\mathbb{T}_{\nu}\times I_0\left(-\epsilon\ln \mathbb{T}_{\nu}\right),
\end{equation}

\noindent where $I_0$ is the Bessel function of the first kind of order zero.

\clearpage

\begin{table}
\begin{center}
\begin{tabular}{|c|c|c|}\hline
$\mathbb{T}_{\nu}$ & $\Delta\mathbb{T}_{\nu}$ & $\frac{\Delta\mathbb{T}_{\nu}}{\mathbb{T}_{\nu}}$ \\ \hline\hline
0.4 & 0.011 & 2.75 \% \\ 
0.6 & 0.009 & 1.5 \% \\\hline
\end{tabular}
\end{center}
\caption{Values of $\Delta\mathbb{T}_{\nu}$ for different values of $\mathbb{T}_{\nu}$ in order to obtain an accuracy of 10 \% on the opacity assuming an uncertainty of 7 \% on the areal mass.}\label{tab1}
\end{table}

\vspace{8cm}

\begin{table}[h]
\begin{center}
\begin{tabular}{|c|c|c|c|c|c|}\hline
Experiment & $\rho L$ & $I_0$ & $\rho$ or $n_e$ & $T$ & $\mathbb{T}$ \\ \hline\hline
Foster \emph{et al.} (1991) \cite{FOSTER91} & & & $\pm$ 20 \% ($\rho$) & & \\
Perry \emph{et al.} (1991) \cite{PERRY91} & & & $\pm$ 35 \% ($\rho$) & $\pm$ 4 \% & $\pm$ 5 \% \\
Springer \emph{et al.} (1992) \cite{SPRINGER92} & & & $\pm$ 0.8 \% ($\rho$) & $\pm$ 5 \% & \\
Da Silva \emph{et al.} (1992) \cite{DASILVA92} & & & & & \\
Eidmann \emph{et al.} (1994) \cite{EIDMANN94} & & $\pm$ 6 \% ($T_{\mathrm{BL}}$) & & & \\
Perry \emph{et al.} (1996) \cite{PERRY96} & $\pm$ 5 \% & & $\pm$ 20 \% ($\rho$) & $\pm$ 5 \% & \\
Merdji \emph{et al.} (1998) \cite{MERDJI98} & & & $\pm$ 35 \% ($\rho$) & & \\
Chenais \emph{et al.} (2000) \cite{CHENAIS00} & & & $\pm$ 50 \% ($\rho$) & $\pm$ 20 \% & \\
Bailey \emph{et al.} (2003) \cite{BAILEY03} & & & $\pm$ 33 \% ($n_e$) & & \\
Renaudin \emph{et al.} (2006) \cite{RENAUDIN06} & $\pm$ 30 \% & & $\pm$ 25 \% ($\rho$) & $\pm$ 2 \% & \\
Loisel \emph{et al.} (2009) \cite{LOISEL09,BLENSKI11a,BLENSKI11b} & & & & & \\
Bailey \emph{et al.} (2007) \cite{BAILEY07} & $\pm$ 25 \% & & $\pm$ 25 \% ($n_e$) & $\pm$ 4 \% & $\pm$ 2 \% \\
                                            &             & &                     &            & in [990-1305 eV]\\
                                            &             & &                     &            & $\pm$ 5 \% \\
                                            &             & &                     &            & in [800-990 eV]\\\hline\hline
\end{tabular}
\caption{Uncertainties mentioned in several publications about absorption-spectroscopy measurements (non-exhaustive list). $T_{\mathrm{BL}}$ represents the effective temperature deduced from the backlighter flux (which is proportional to the fourth power of $T_{\mathrm{BL}}$ according to Stefan's law).}\label{tab2}
\end{center}
\end{table}

\begin{table}
\begin{center}
\begin{tabular}{|c|c|c|}\hline
 Defect & Areal mass & Modified transmission \\ \hline\hline
Wedge & $\rho L(x)=\rho_1L_1\left(1-\epsilon\frac{x}{a}\right)$ & $\tilde{\mathbb{T}}_{\nu}=\frac{\left(1-\epsilon/2\right)}{\epsilon\ln\left(\mathbb{T}_{\nu}\right)}\left[\mathbb{T}_{\nu}\right]^{\frac{1}{1-\epsilon/2}}\left[1-\mathbb{T}_{\nu}^{-\frac{\epsilon}{1-\epsilon/2}}\right]$ \\ 
Bulge & $\rho L(x)=\rho_1L_1\left[-\frac{4\epsilon}{a^2}x(x-a)+1\right]$ & $\tilde{\mathbb{T}}_{\nu}=\sqrt{\frac{1+2\epsilon/3}{-\epsilon\ln\mathbb{T}_{\nu}}}~\left[\mathbb{T}_{\nu}\right]^{\frac{1}{1+2\epsilon/3}}~D\left(\sqrt{\frac{-\epsilon\ln\mathbb{T}_{\nu}}{1+2\epsilon/3}}\right)$ \\
Hollow & $\rho L(x)=\rho_1L_1\left[\frac{4\epsilon}{a^2}x(x-a)+1\right]$ & $\tilde{\mathbb{T}}_{\nu}=\frac{\sqrt{\pi}}{2}\left[\mathbb{T}_{\nu}\right]^{\frac{1-\epsilon}{1-2\epsilon/3}}~\sqrt{\frac{1-2\epsilon/3}{-\epsilon\ln\mathbb{T}_{\nu}}}~\mathrm{Erf}\left[\sqrt{\frac{-\epsilon\ln\mathbb{T}_{\nu}}{1-2\epsilon/3}}\right]$\\
Holes & $\rho_0L_0 \rightarrow\rho_1L_1(1-\epsilon)$ & $\tilde{\mathbb{T}}_{\nu}=(1-\epsilon)\left[\mathbb{T}_{\nu}\right]^{\frac{1}{1-\epsilon}}+\epsilon$\\
Modulations & $\rho L(x)=\rho_0L_0\left[1+\epsilon\cos(x)\right]$ & $
\tilde{\mathbb{T}_{\nu}}=\mathbb{T}_{\nu}\times I_0\left(-\epsilon\ln \mathbb{T}_{\nu}\right)$\\\hline
\end{tabular}
\end{center}
\caption{Modified transmission $\tilde{\mathbb{T}}_{\nu}$ as a function of the ``unperturbed'' transmission $\mathbb{T}_{\nu}$ in the case of the different areal-mass defects considered in the present paper. $D(x)=e^{-x^2}\int_0^x e^{t^2}dt$ represents Dawson's function, $\mathrm{Erf}(x)=\frac{2}{\sqrt{\pi}}\int_0^xe^{-t^2}dt$ is the usual error function, and $I_0(x)$ is the Bessel function of the first kind of order zero. $a$ is the areal size of the sample and $\epsilon$ quantifies the amplitude of the perturbation.}\label{tab3}
\end{table}

\begin{table}
\begin{center}
\begin{tabular}{|c|c|}\hline
Kind of irregularity & Value of $\epsilon$ (\% of distortion) \\ \hline\hline
Wedge & 0.4 (40 \%) \\ 
Bulge & 0.4 (80 \%) \\
Hollow & 0.4 (60 \%) \\
Holes & 0.05 (5 \%) \\ 
Modulations & 0.2 (20 \%) \\\hline
\end{tabular}
\end{center}
\caption{Values of $\epsilon$ required to obtain an uncertainty of 7 \% on the areal mass (estimated from Rutherford back-scattering), as in the NIF experiment \cite{HEETER17,PERRY17,HEETER18}.}\label{tab4}
\end{table}

\clearpage

\begin{figure}[ht]
\vspace{1cm}
\begin{center}
\includegraphics[width=11cm]{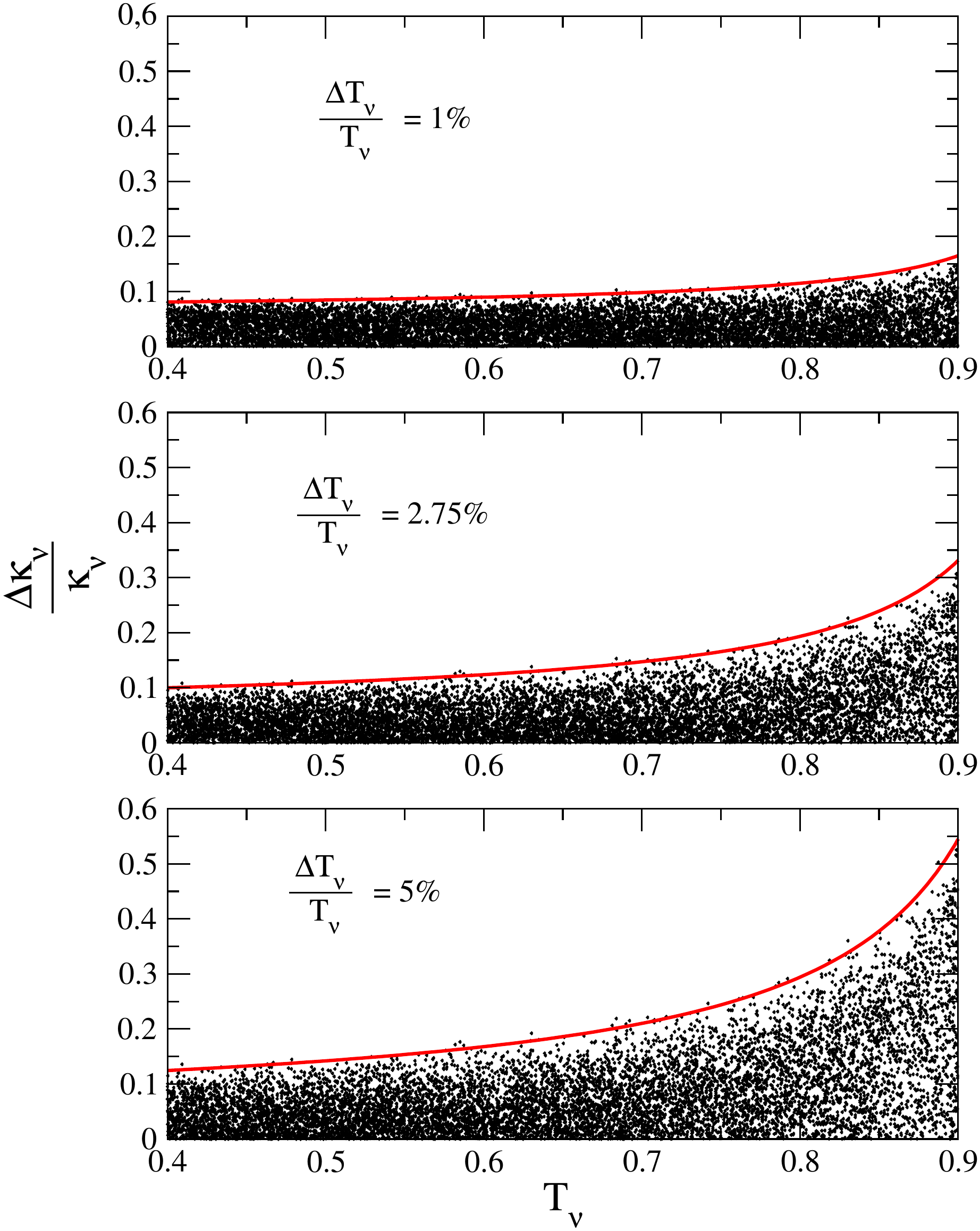}
\end{center}
\caption{(Color online) Monte Carlo simulation of the propagation of uncertainties. In the three cases, the points correspond to values of $\Delta\kappa_{\nu}/\kappa_{\nu}=\left|\kappa_i-\kappa_0\right|/\kappa_0$, i=1,10000 and $\Delta(\rho L)/(\rho L)$=0.07. The first case (top) corresponds to $\Delta \mathbb{T}_{\nu}/\mathbb{T}_{\nu}$=1 \%, the second case (middle) to $\Delta \mathbb{T}_{\nu}/\mathbb{T}_{\nu}$=2.75 \% and the last case (bottom) to $\Delta \mathbb{T}_{\nu}/\mathbb{T}_{\nu}$=5 \%. As mensioned in the text, the results do not depend on $\kappa_0$.}\label{figure_new}
\vspace{1cm}
\end{figure}

\clearpage

\begin{figure}[ht]
\vspace{1cm}
\begin{center}
\includegraphics[width=11cm]{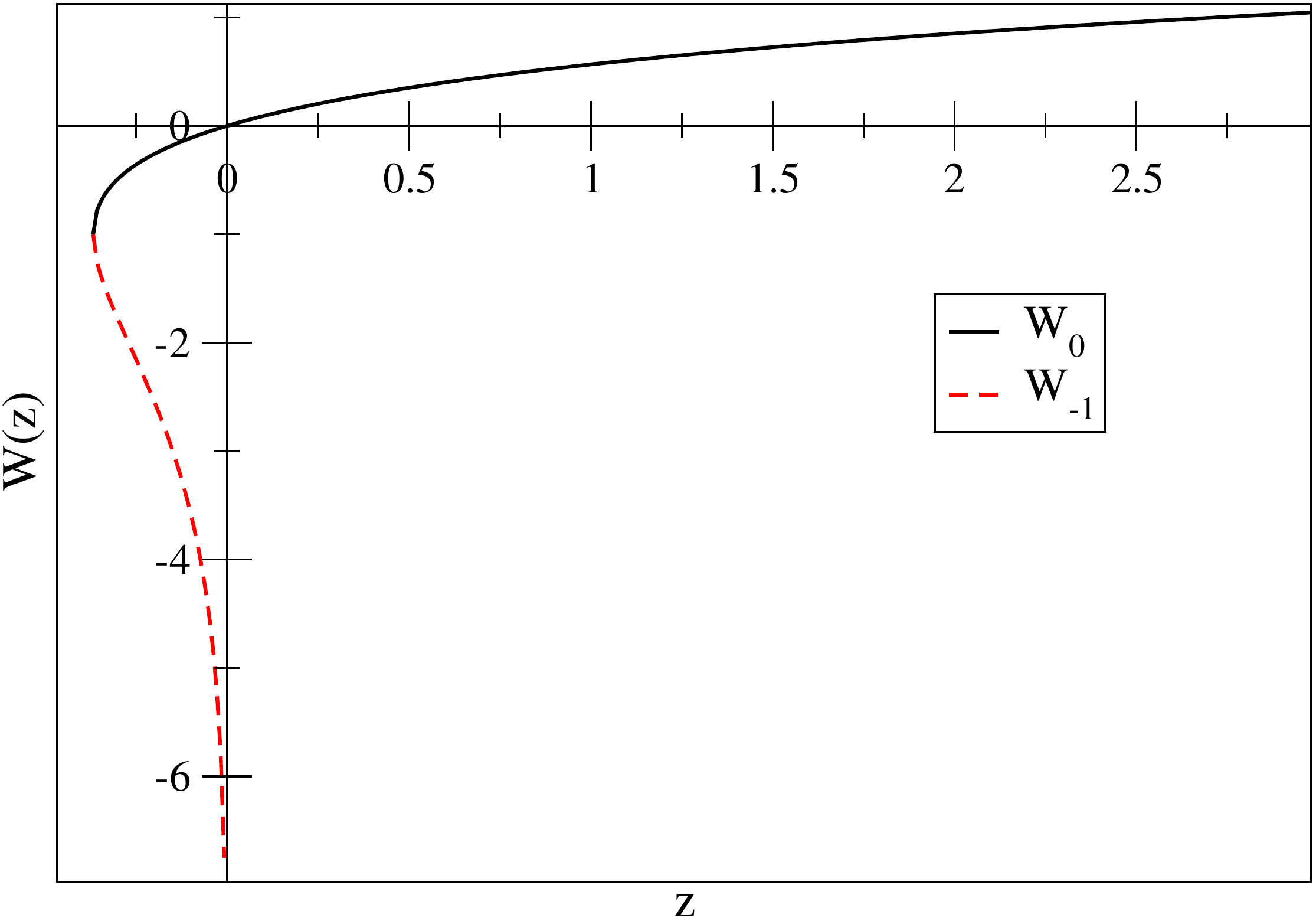}
\end{center}
\caption{(Color online) The two real branches of Lambert function: $W_0$ and $W_{-1}$.}\label{fig_lambert}
\vspace{1cm}
\end{figure}

\begin{figure}[ht]
\vspace{1cm}
\begin{center}
\includegraphics[width=11cm]{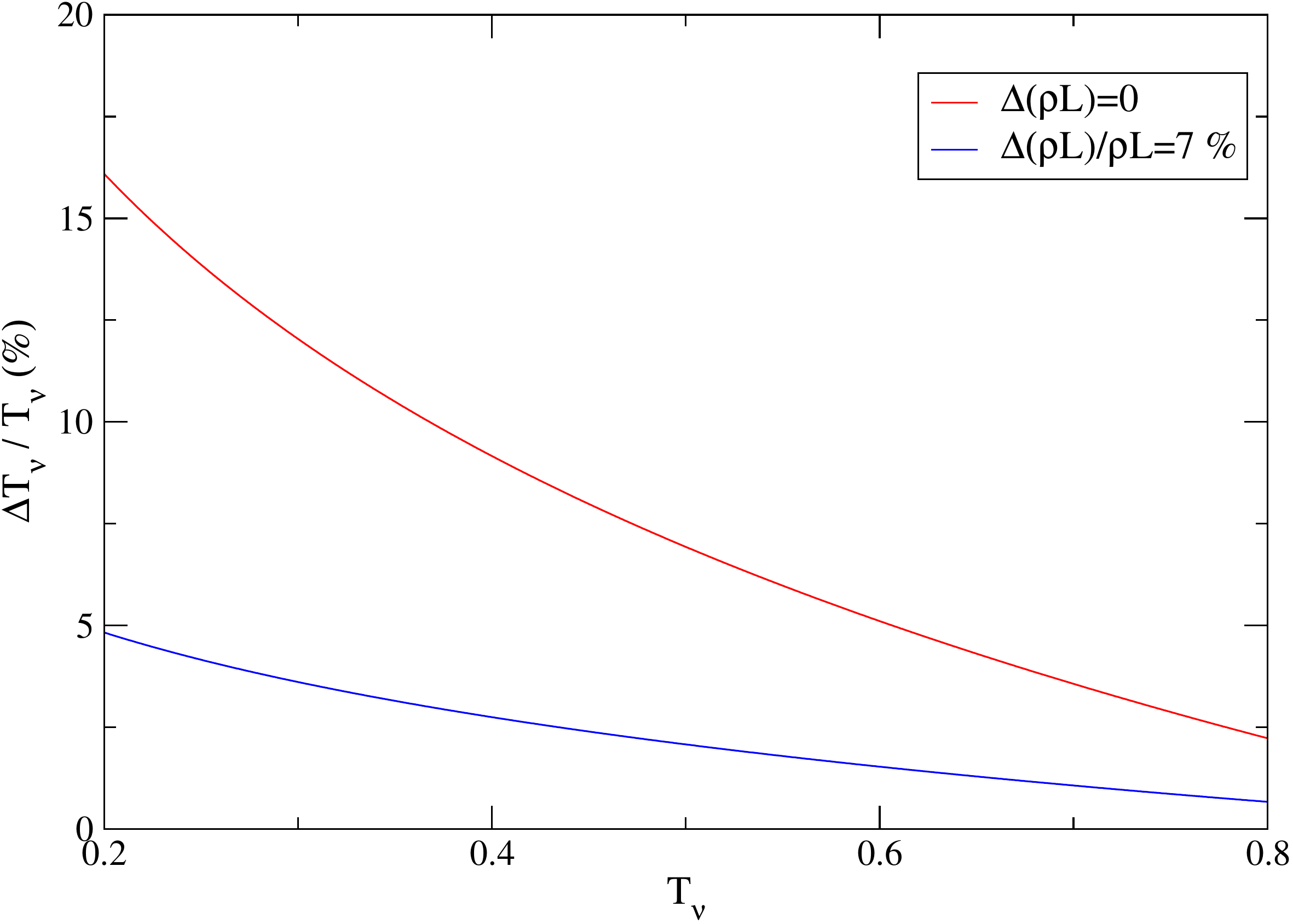}
\end{center}
\caption{(Color online) Value of $\Delta\mathbb{T}_{\nu}/\mathbb{T}_{\nu}$ as a function of $\mathbb{T}_{\nu}$ for a required precision on opacity $\Delta\kappa_{\nu}/\kappa_{\nu}=10 \%$ and two cases: $\Delta(\rho L)=0$ (red curve, see Eq. (\ref{lim1})) and $\Delta(\rho L)/(\rho L)=7 \%$ (blue curve, see Eq. (\ref{lim2})).}\label{figt1t2}
\vspace{1cm}
\end{figure}

\begin{figure}[ht]
\vspace{1cm}
\begin{center}
\includegraphics[width=11cm]{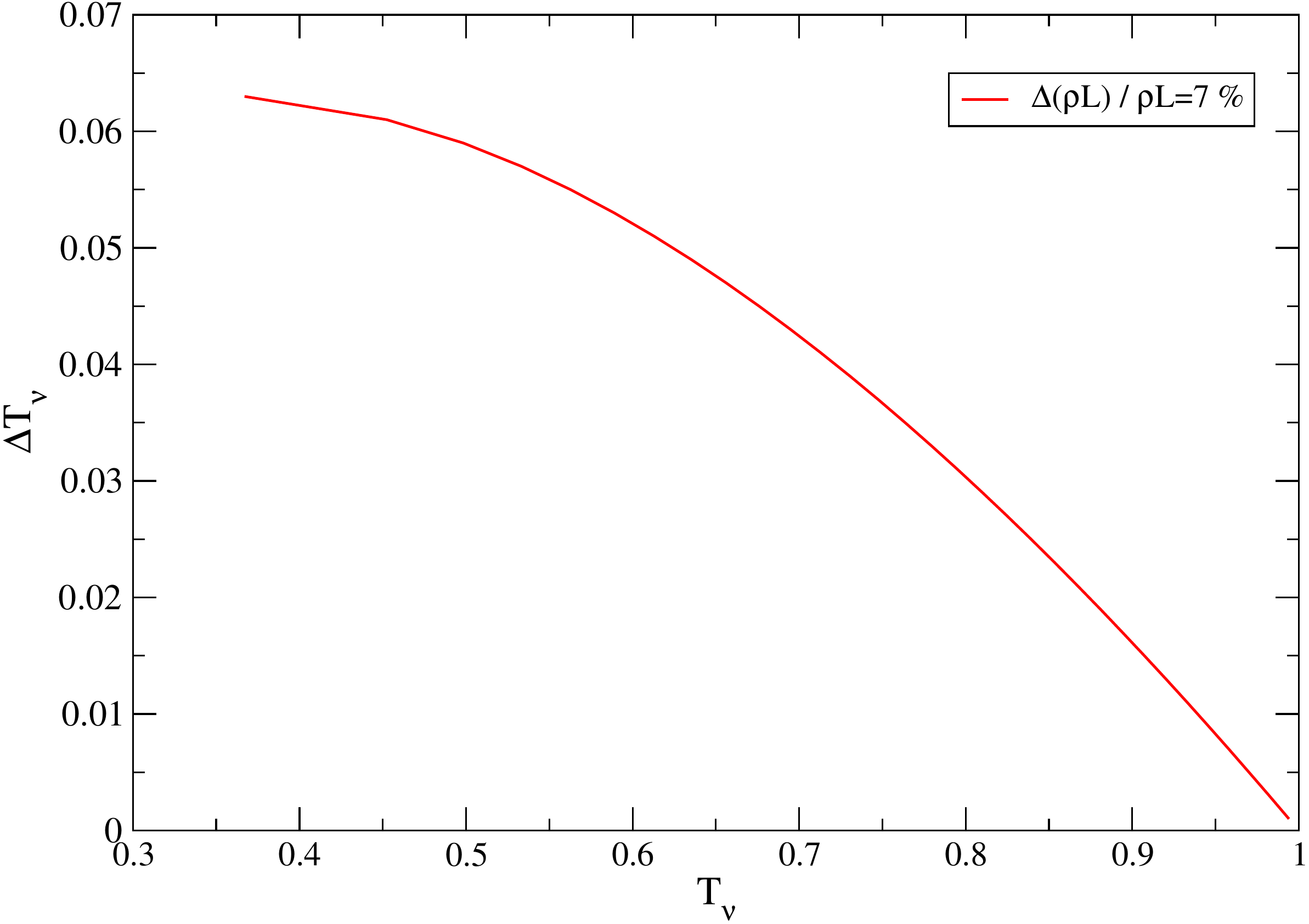}
\end{center}
\caption{(Color online) Value of $\Delta\mathbb{T}_{\nu}$ as a function of $\mathbb{T}_{\nu}$ (see Eq. (\ref{deltatnu})).}\label{fig_deltatnu}
\vspace{1cm}
\end{figure}

\begin{figure}[ht]
\vspace{1cm}
\begin{center}
\includegraphics[width=11cm]{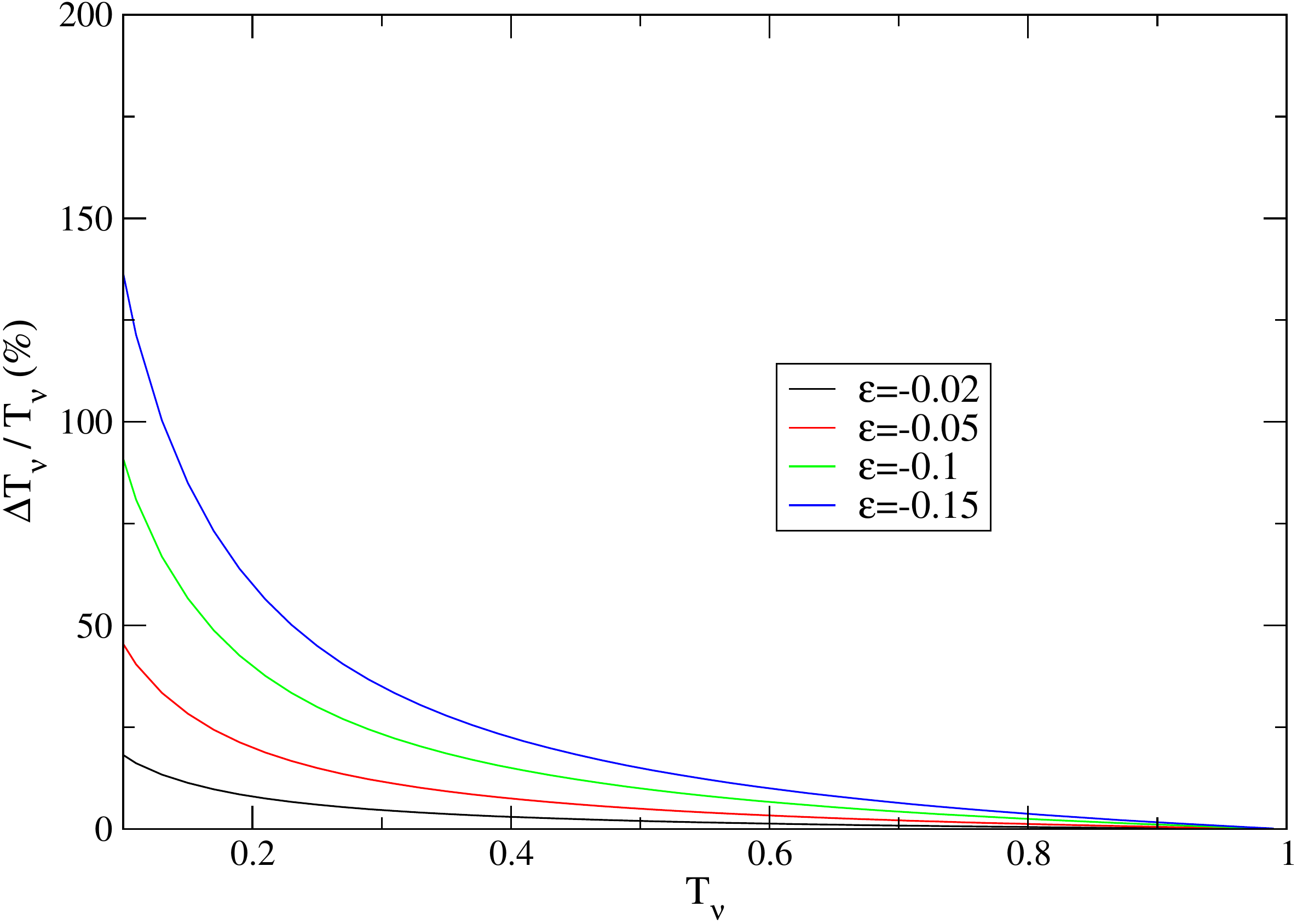}
\end{center}
\caption{(Color online) Impact of background radiation on $\Delta\mathbb{T}_{\nu}/\mathbb{T}_{\nu}$ for different values of $\epsilon=\delta/I_0$.}\label{fig_back_1}
\vspace{1cm}
\end{figure}

\begin{figure}[ht]
\vspace{1cm}
\begin{center}
\includegraphics[width=11cm]{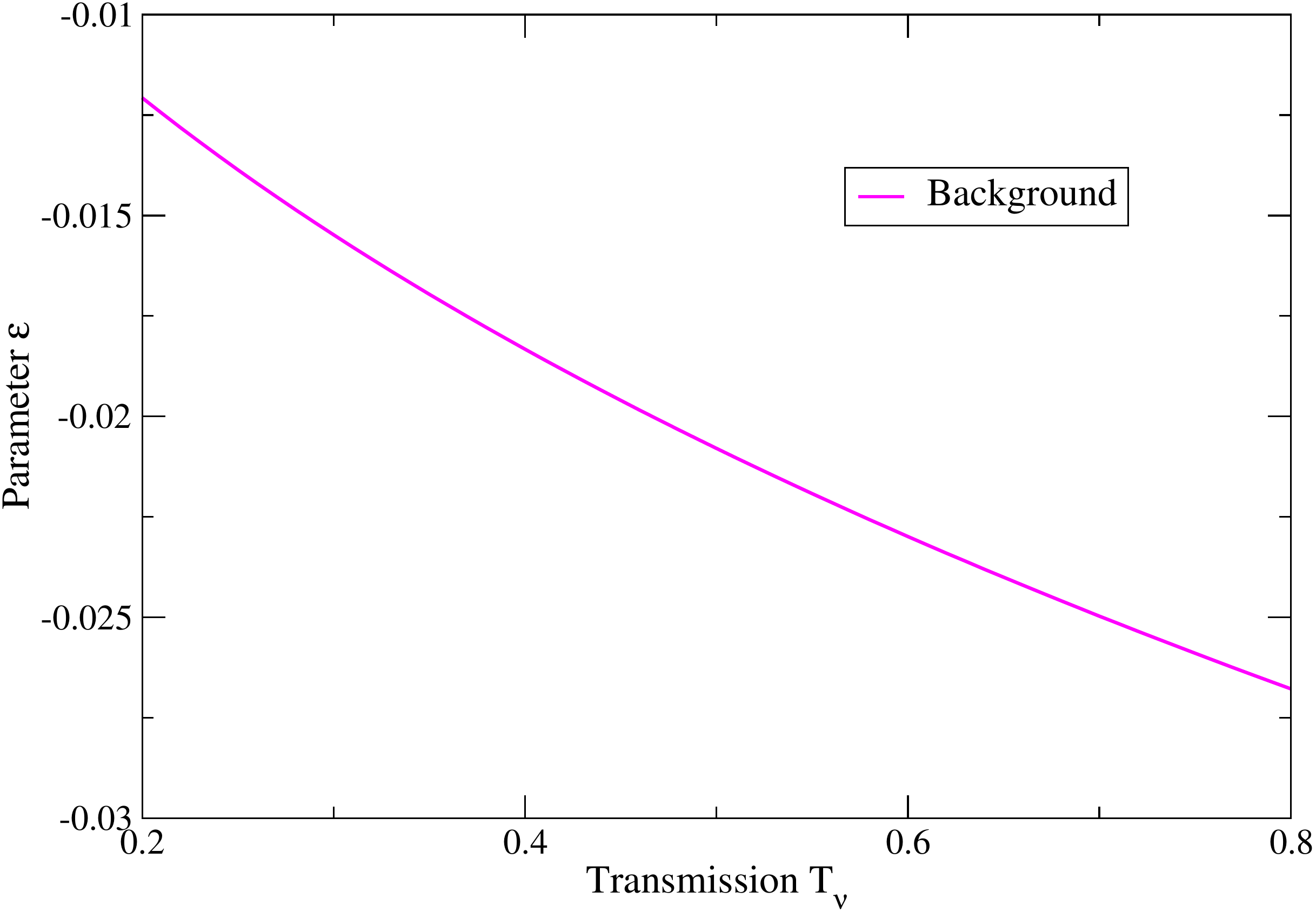}
\end{center}
\caption{(Color online) Variation of parameter $\epsilon$ as a function of transmission in the case of a background perturbation. One has $\Delta\kappa_{\nu}/\kappa_{\nu}=10 \%$ and $\Delta(\rho L)/(\rho L)=7 \%$.}\label{data6-background}
\vspace{1cm}
\end{figure}

\begin{figure}[ht]
\vspace{1cm}
\begin{center}
\includegraphics[width=11cm]{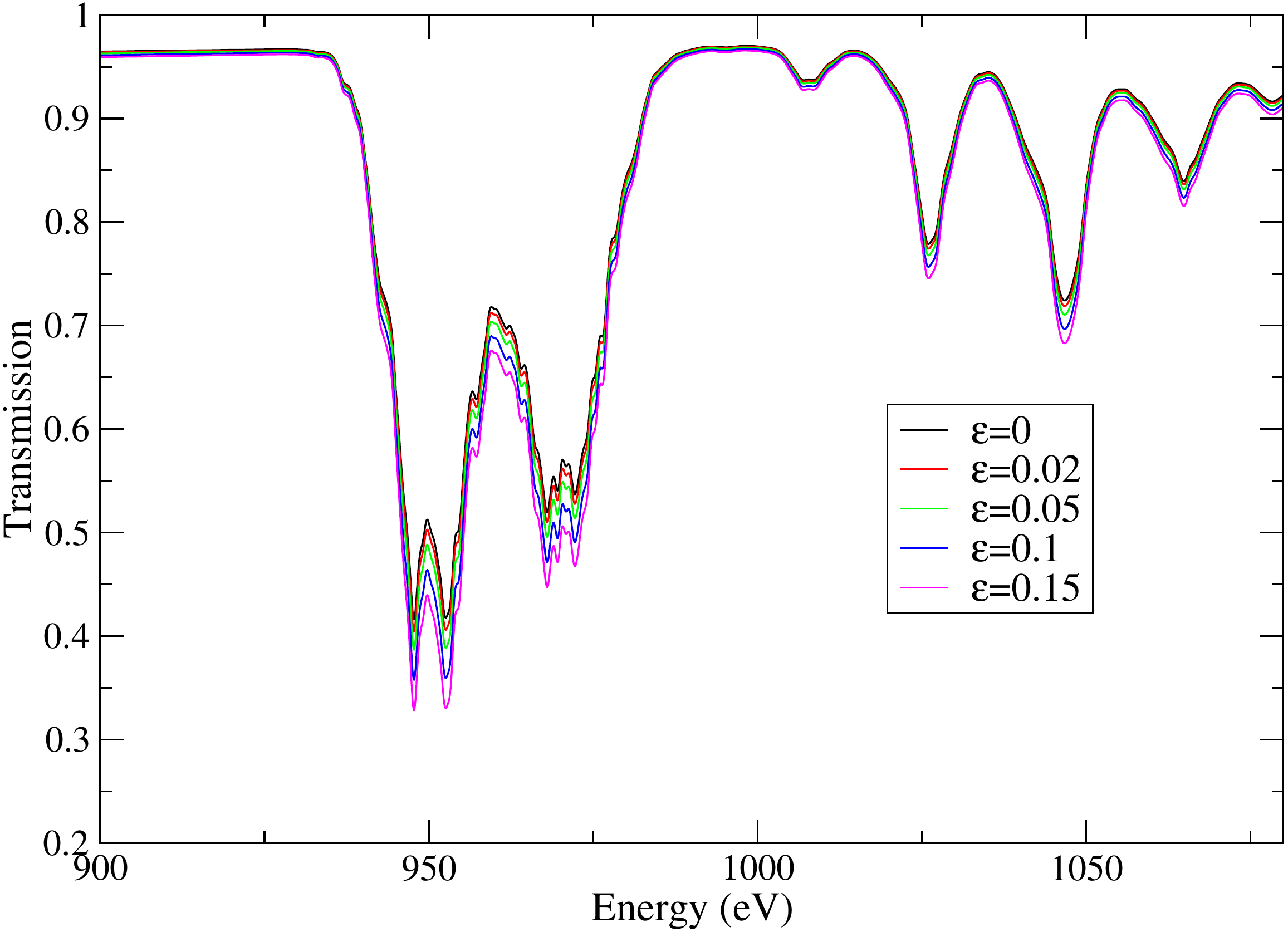}
\end{center}
\caption{(Color online) Impact of background radiation on the transmission of copper at $T$=18 eV, $\rho$=0.01 g/cm$^3$ and a resolving power of $R=E/\Delta E$=1000 ($E$ is the photon energy) for different values of $\epsilon=\delta/I_0$.}\label{fig_back_2}
\vspace{1cm}
\end{figure}

\begin{figure*}
\vspace{1cm}
\begin{center}
\includegraphics[width=11cm]{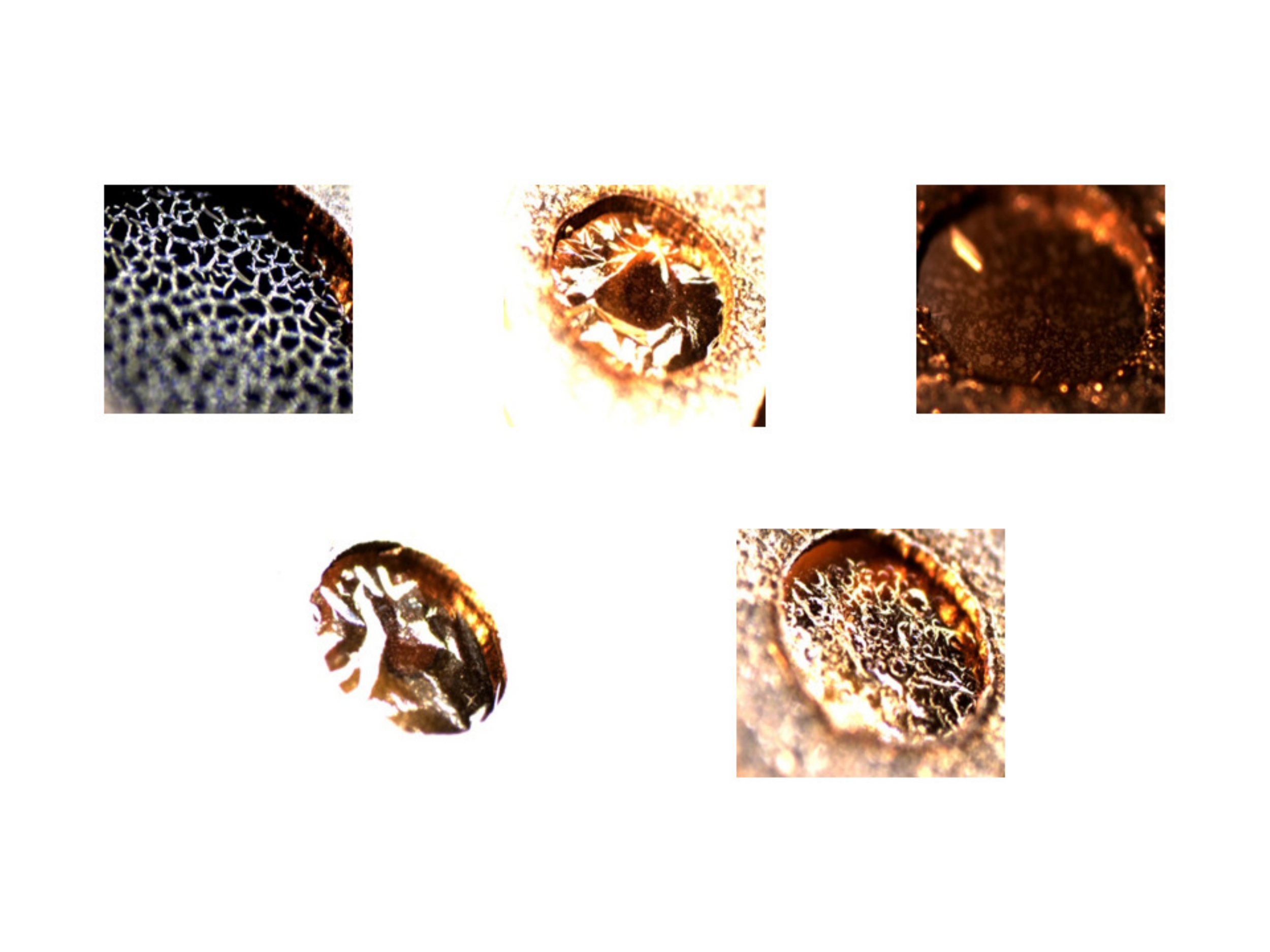}
\end{center}
\caption{(Color online) Microscopy views of a few samples from a recent (2017) absorption-spectroscopy campaign at LULI laboratory \cite{POIRIER18}.}\label{defects}
\vspace{1cm}
\end{figure*}

\begin{figure*}
\vspace{1cm}
\begin{center}
\includegraphics[width=11cm]{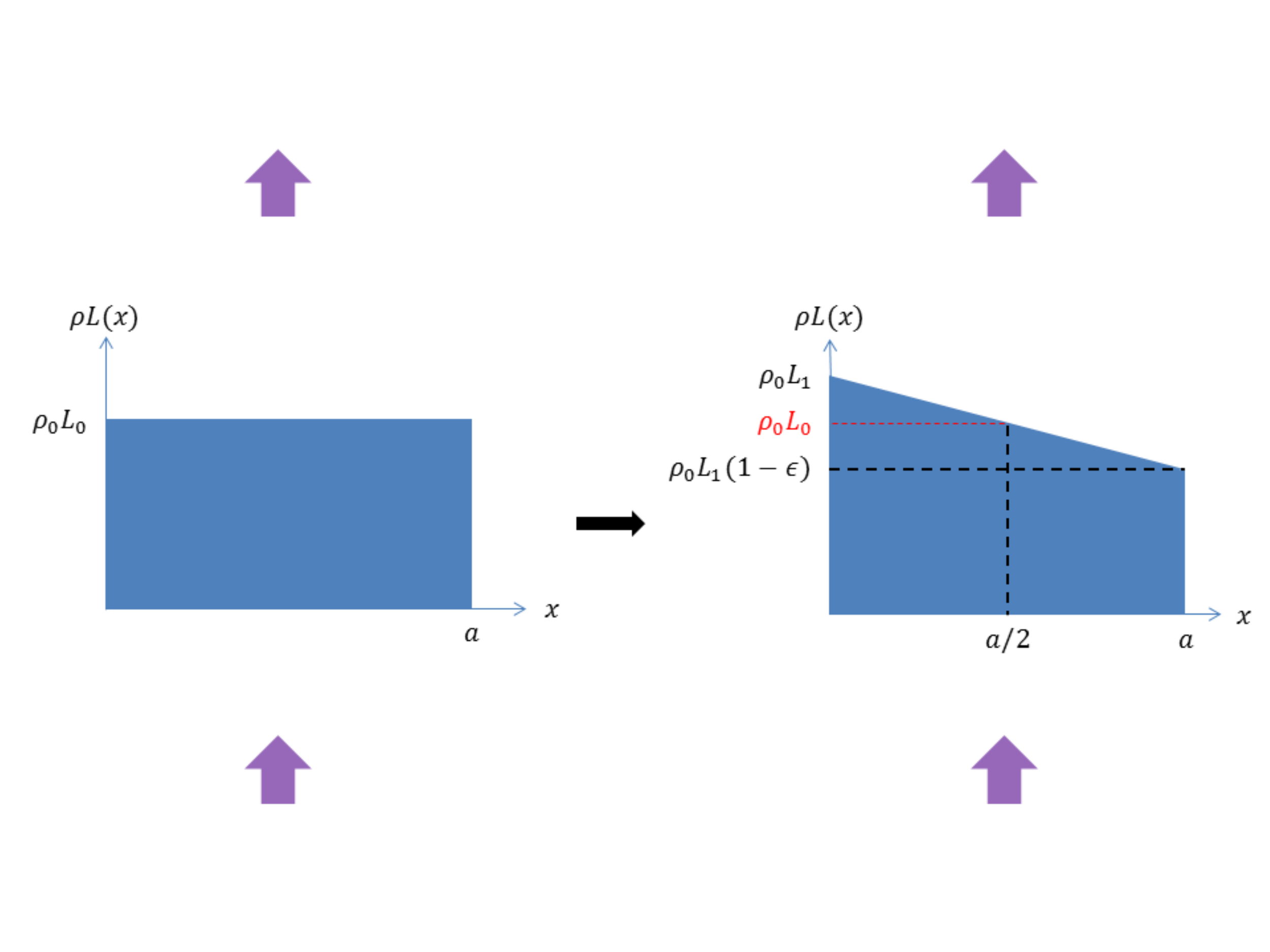}
\end{center}
\caption{(Color online) Wedge-shape distortion of the sample. The lowest purple arrows corresponds to $I_0$ and the highest to $I_{\nu}$. For the sample on the left $I_{\nu}=\mathbb{T}_{\nu}I_0$ and for the sample on the right $I_{\nu}=\tilde{\mathbb{T}_{\nu}}I_0$.} \label{wedge}
\vspace{1cm}
\end{figure*}

\begin{figure}[ht]
\vspace{1cm}
\begin{center}
\includegraphics[width=11cm]{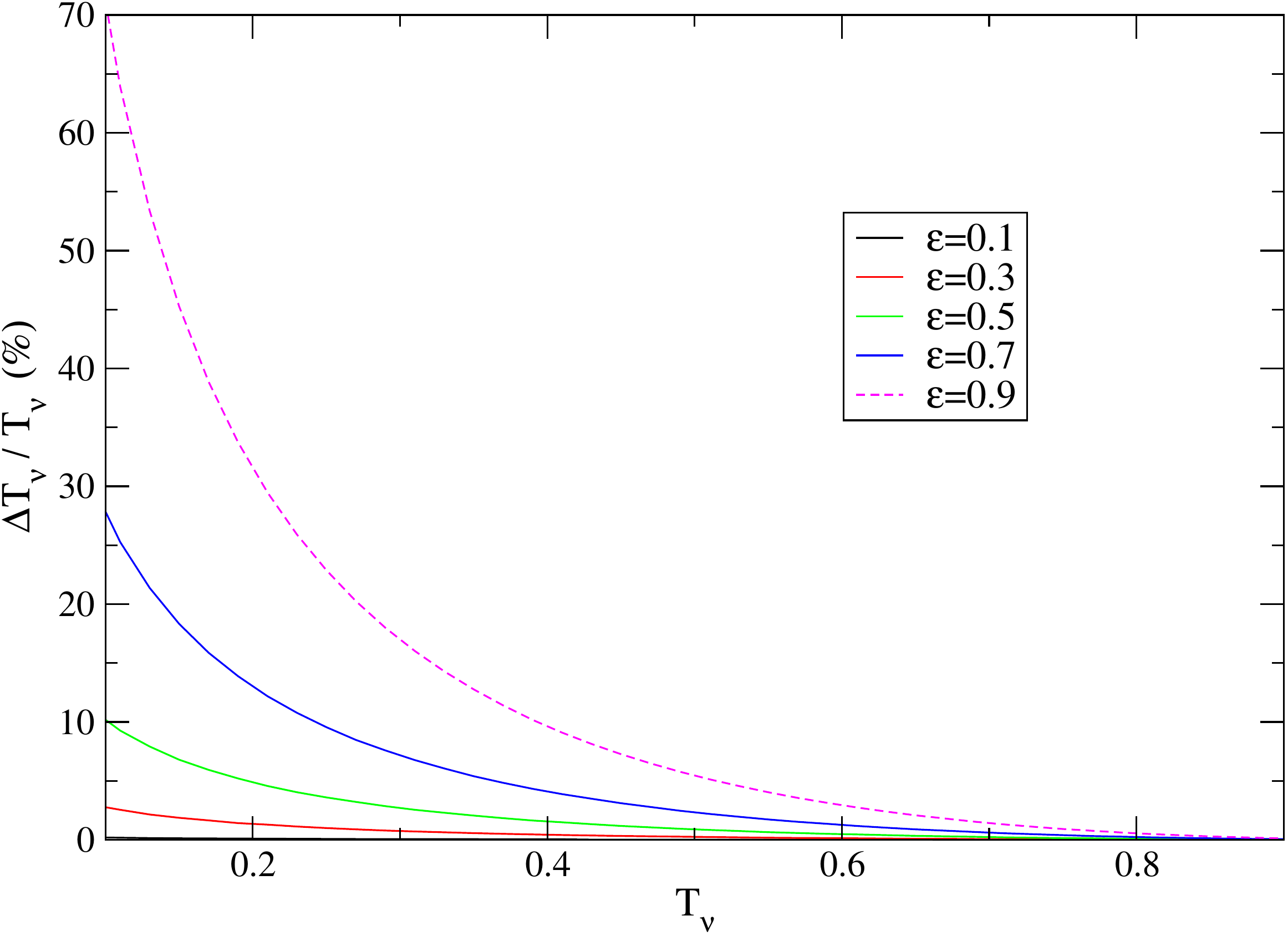}
\end{center}
\caption{(Color online) Impact of a wedge-shape distortion of the sample on $\Delta\mathbb{T}_{\nu}/\mathbb{T}_{\nu}$ for different values of $\epsilon$.}\label{fig_wedge_1}
\vspace{1cm}
\end{figure}

\begin{figure}[ht]
\vspace{1cm}
\begin{center}
\includegraphics[width=11cm]{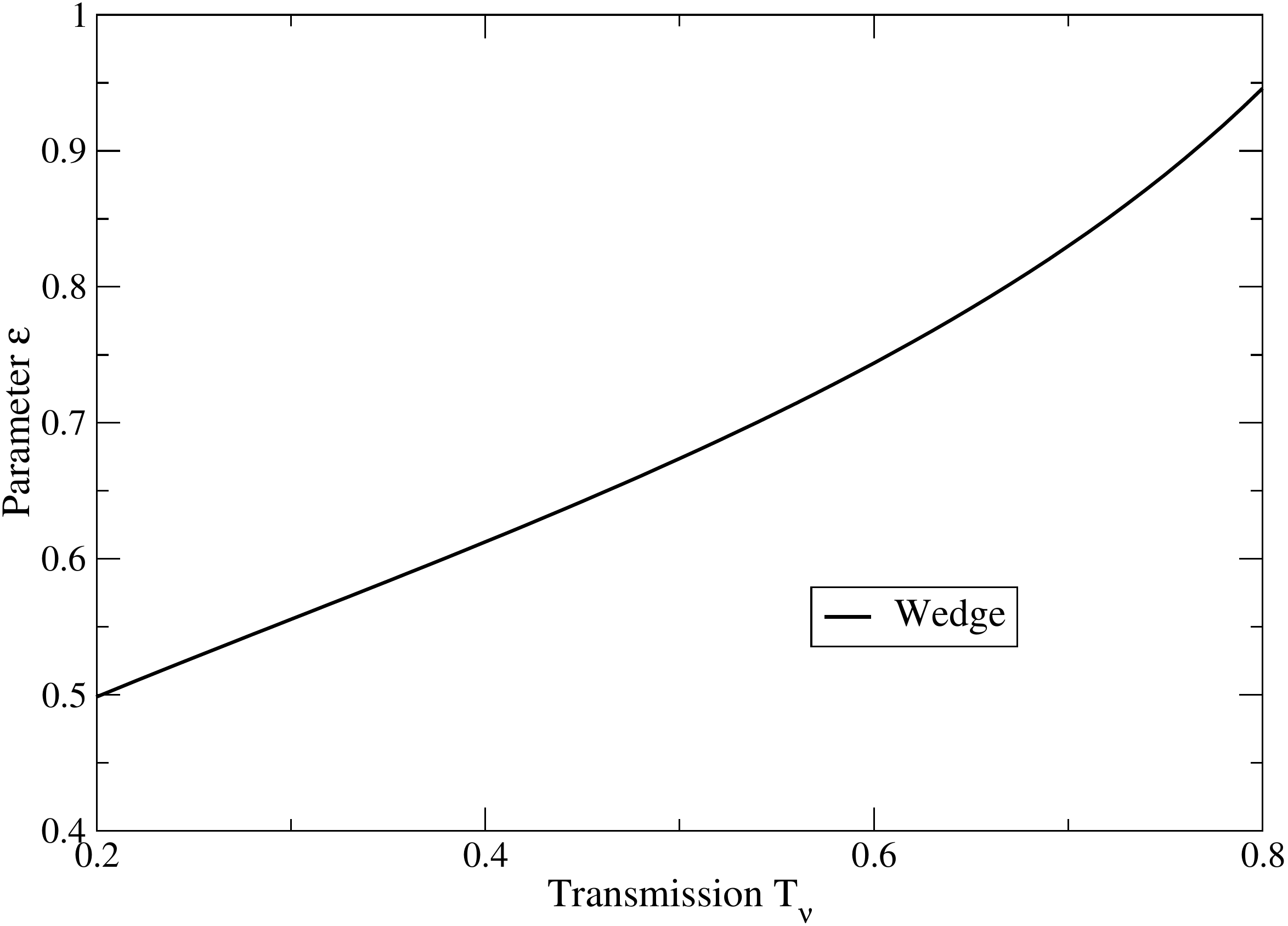}
\end{center}
\caption{(Color online) Variation of parameter $\epsilon$ as a function of transmission in the case of a wedge-shape distortion of the sample. One has $\Delta\kappa_{\nu}/\kappa_{\nu}=10 \%$ and $\Delta(\rho L)/(\rho L)=7 \%$.}\label{data1-wedge}
\vspace{1cm}
\end{figure}

\begin{figure}[ht]
\vspace{1cm}
\begin{center}
\includegraphics[width=11cm]{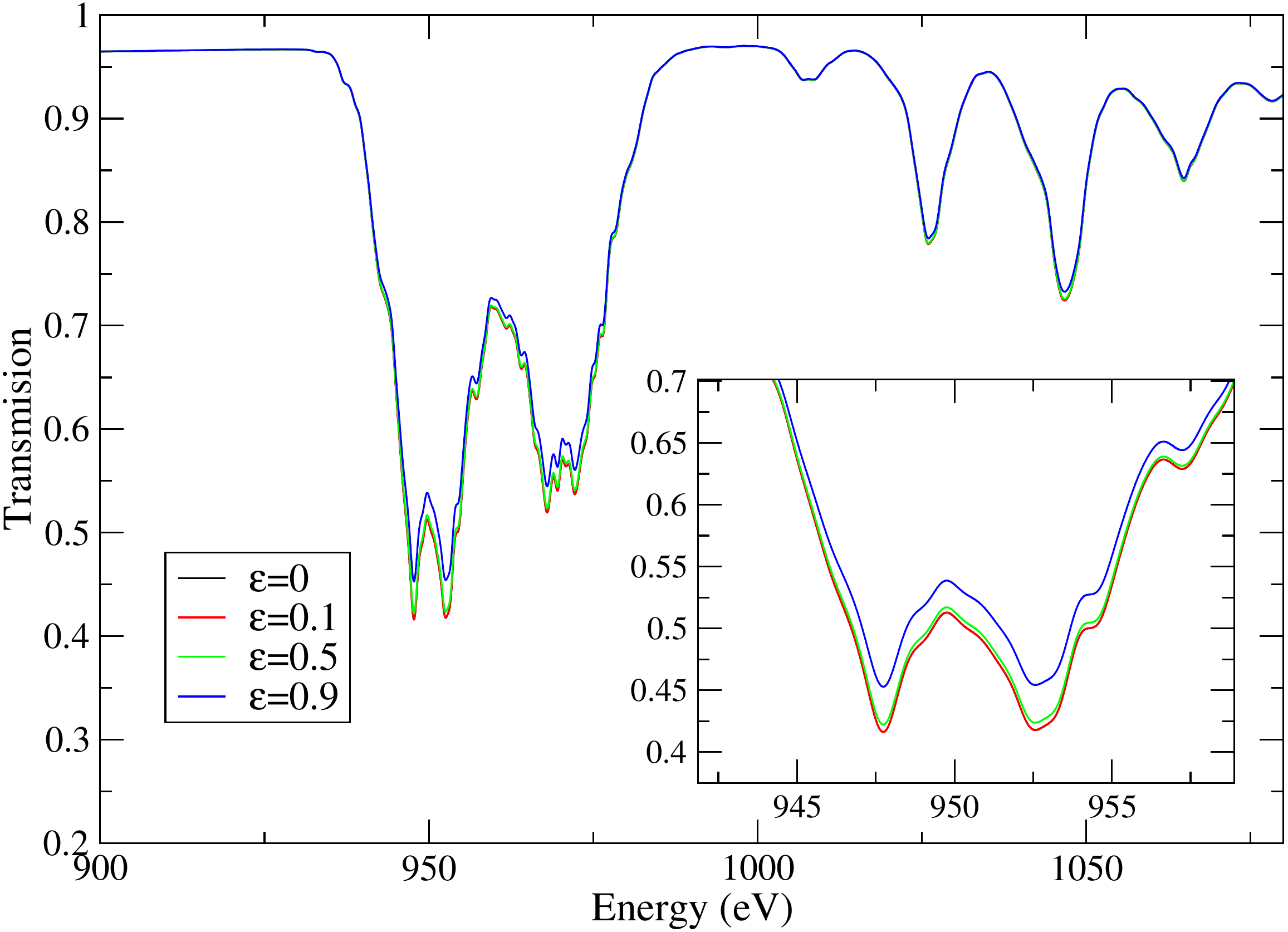}
\end{center}
\caption{(Color online) Impact of a wedge-shape distortion of the sample on the transmission of copper at $T$=18 eV, $\rho$=0.01 g/cm$^3$ and a resolving power of $R=E/\Delta E$=1000 ($E$ is the photon energy) for different values of $\epsilon$.}\label{fig_wedge_2}
\vspace{1cm}
\end{figure}
 
\begin{figure*}
\vspace{1cm}
\begin{center}
\includegraphics[width=11cm]{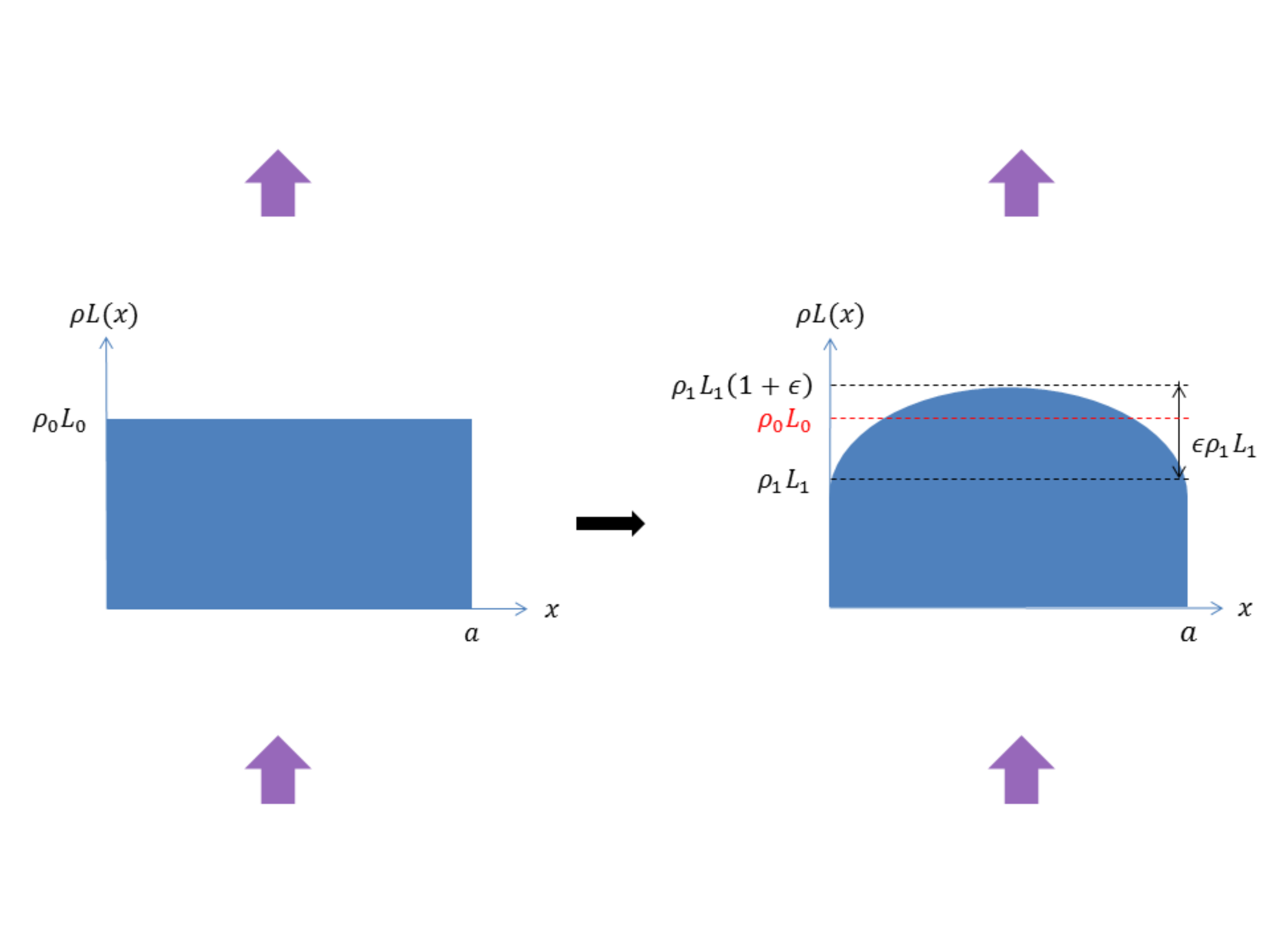}
\end{center}
\caption{(Color online) Concave distortion of the sample (bulge). The lowest purple arrows corresponds to $I_0$ and the highest to $I_{\nu}$. For the sample on the left $I_{\nu}=\mathbb{T}_{\nu}I_0$ and for the sample on the right $I_{\nu}=\tilde{\mathbb{T}_{\nu}}I_0$.}\label{bulge}
\vspace{1cm}
\end{figure*}

\begin{figure}[ht]
\vspace{1cm}
\begin{center}
\includegraphics[width=11cm]{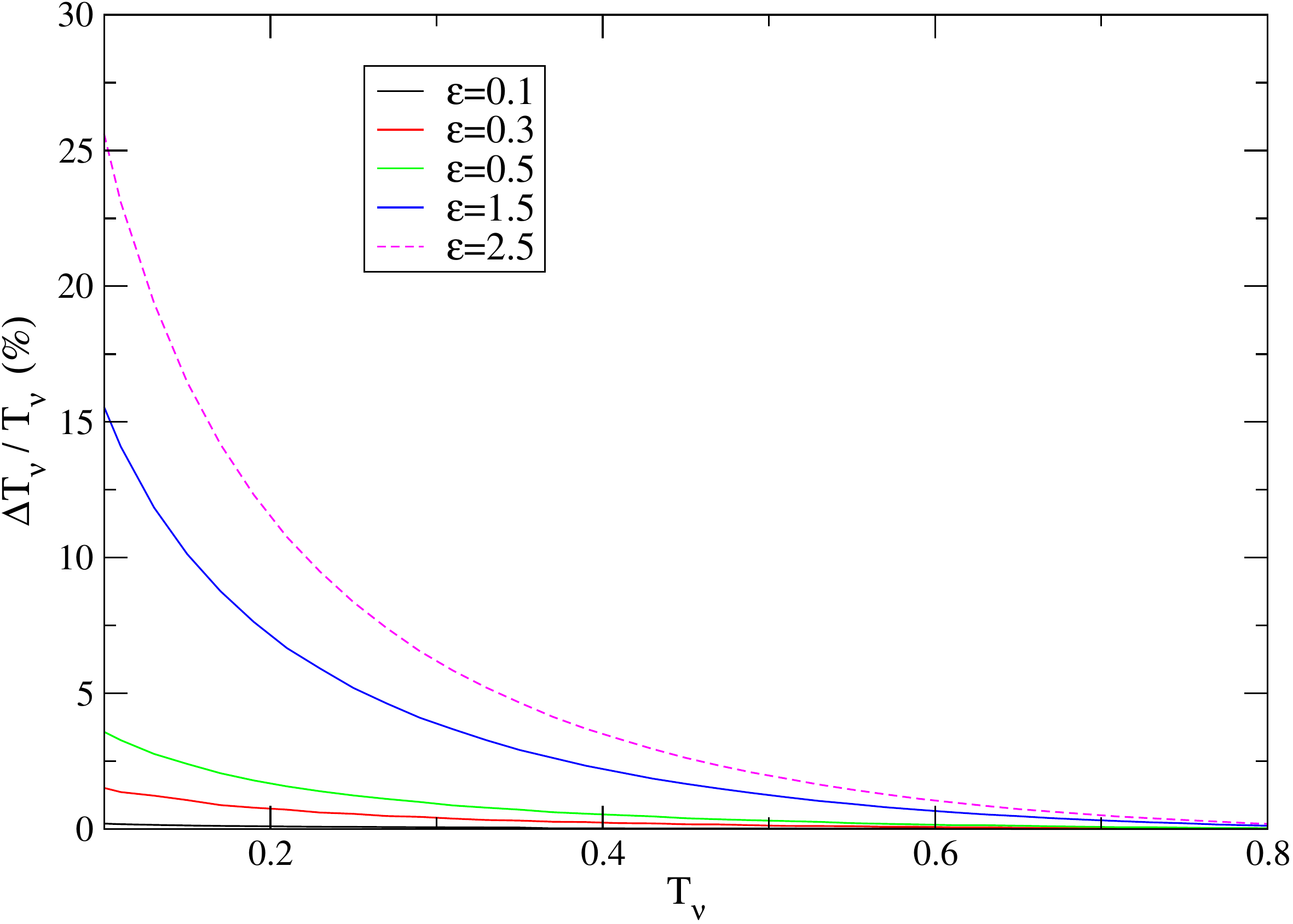}
\end{center}
\caption{(Color online) Impact of a concave distortion of the sample (bulge) on $\Delta\mathbb{T}_{\nu}/\mathbb{T}_{\nu}$ for different values of $\epsilon$.}\label{fig_bulge_1}
\vspace{1cm}
\end{figure}

\begin{figure}[ht]
\vspace{1cm}
\begin{center}
\includegraphics[width=11cm]{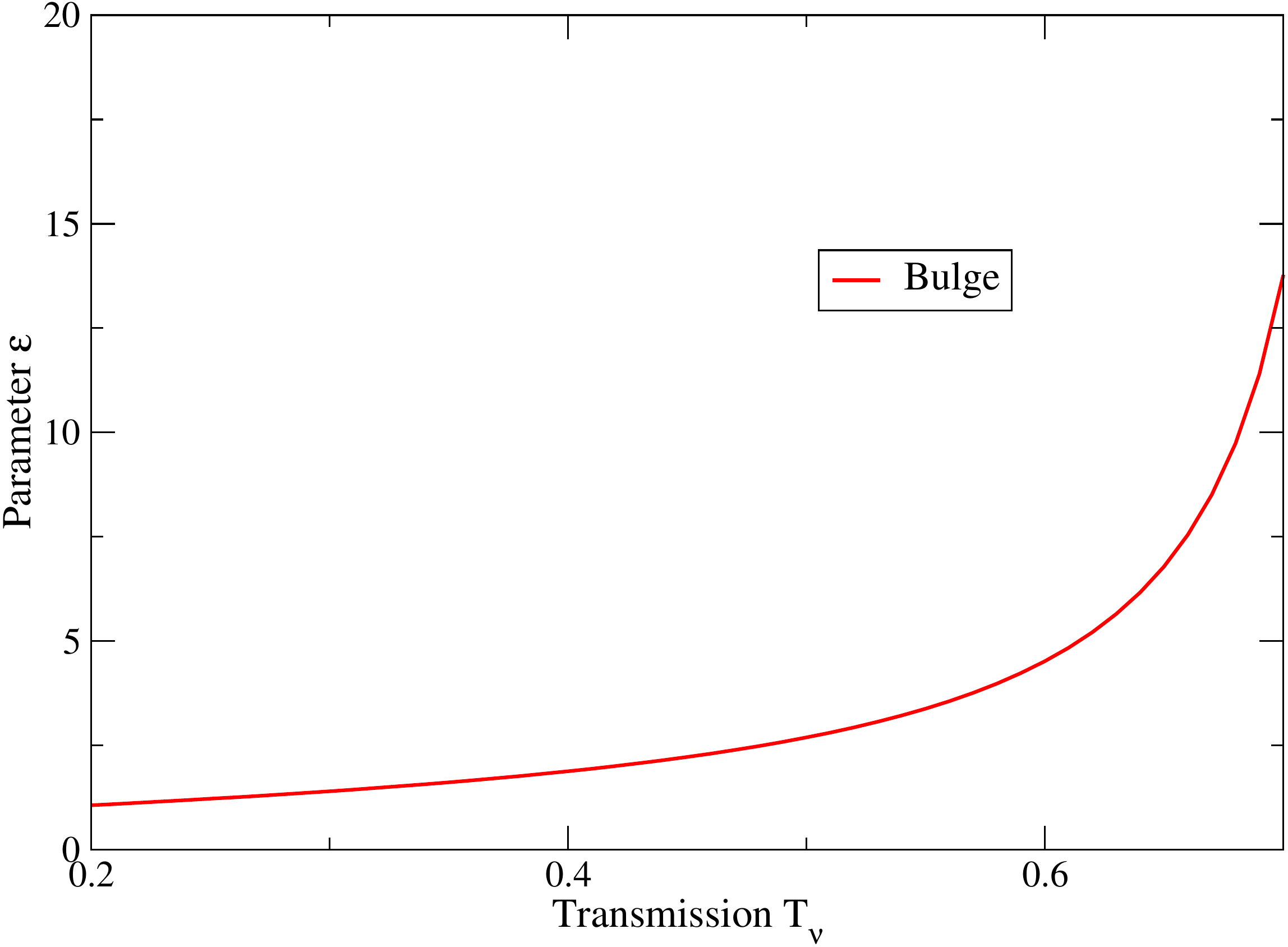}
\end{center}
\caption{(Color online) Variation of parameter $\epsilon$ as a function of transmission in the case of a bulge-shape distortion of the sample. One has $\Delta\kappa_{\nu}/\kappa_{\nu}=10 \%$ and $\Delta(\rho L)/(\rho L)=7 \%$.}\label{data2-bulge}
\vspace{1cm}
\end{figure}

\clearpage

\begin{figure}[ht]
\vspace{1cm}
\begin{center}
\includegraphics[width=11cm]{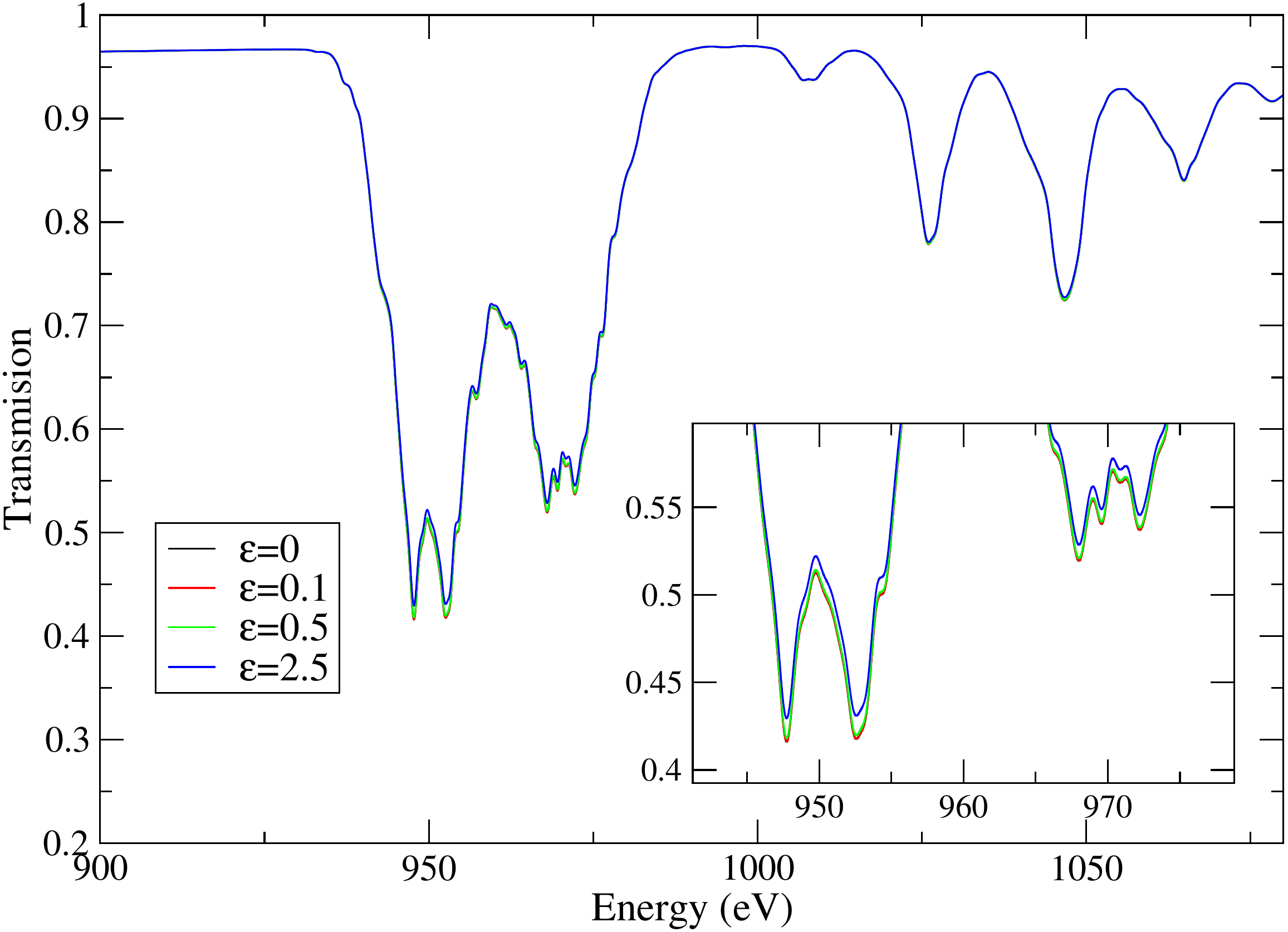}
\end{center}
\caption{(Color online) Impact of a concave distortion of the sample (bulge) on the transmission of copper at $T$=18 eV, $\rho$=0.01 g/cm$^3$ and a resolving power of $R=E/\Delta E$=1000 ($E$ is the photon energy) for different values of $\epsilon$.}\label{fig_bulge_2}
\vspace{1cm}
\end{figure}

\begin{figure*}
\vspace{1cm}
\begin{center}
\includegraphics[width=11cm]{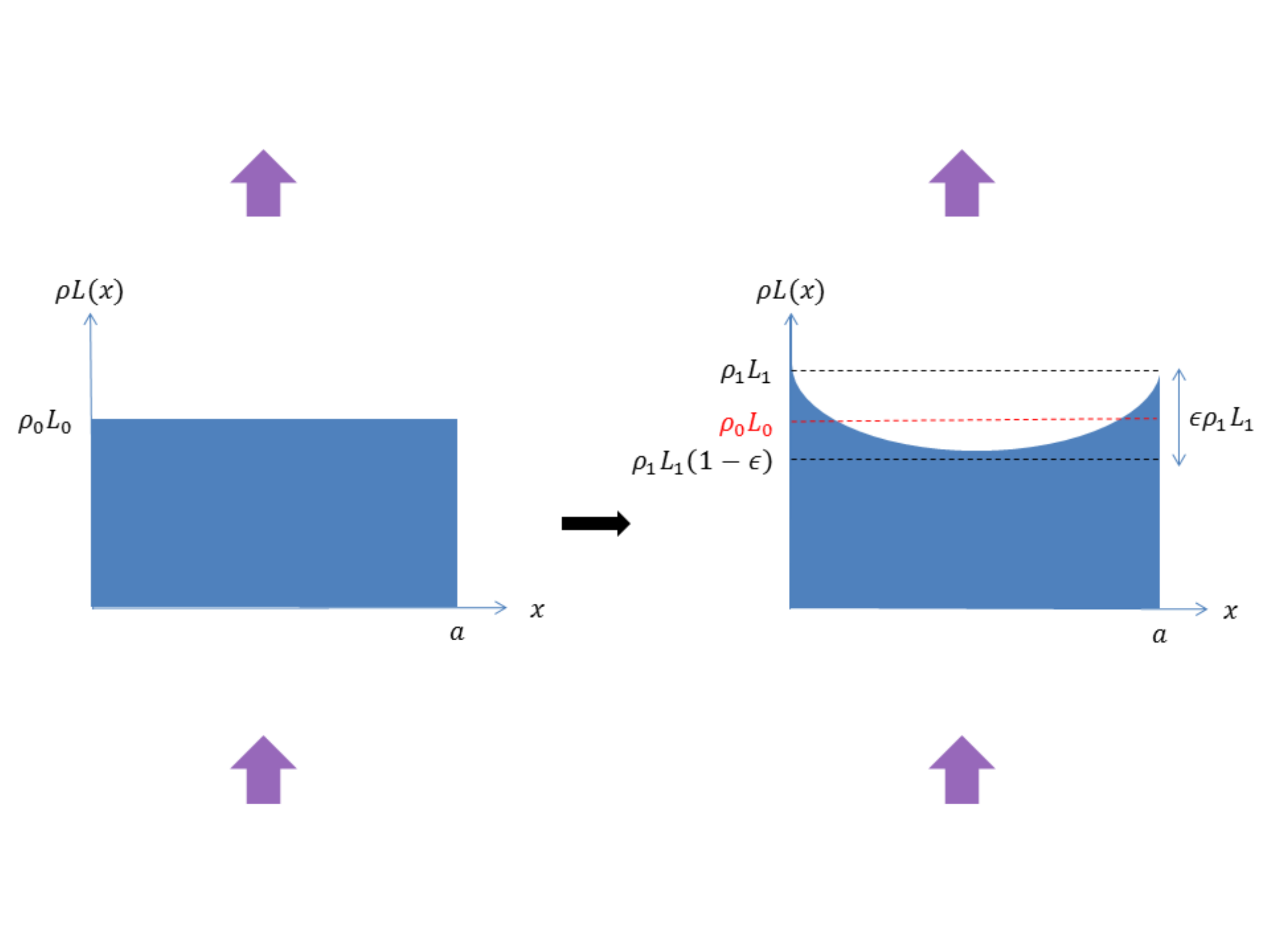}
\end{center}
\caption{(Color online) Convex distortion of the sample (hollow). The lowest purple arrows corresponds to $I_0$ and the highest to $I_{\nu}$. For the sample on the left $I_{\nu}=\mathbb{T}_{\nu}I_0$ and for the sample on the right $I_{\nu}=\tilde{\mathbb{T}_{\nu}}I_0$.}\label{hollow}
\vspace{1cm}
\end{figure*}

\begin{figure}[ht]
\vspace{1cm}
\begin{center}
\includegraphics[width=11cm]{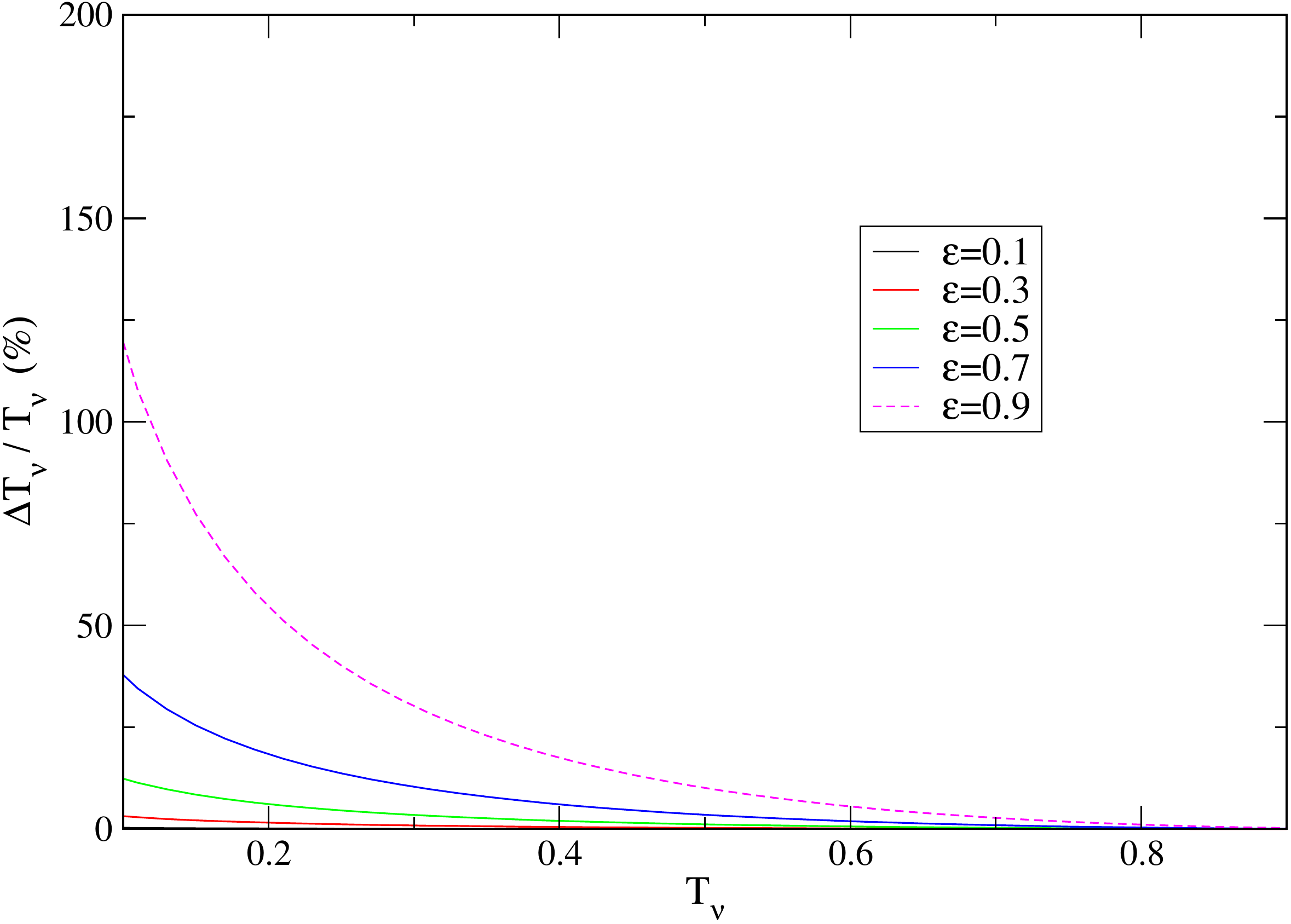}
\end{center}
\caption{(Color online) Impact of a convex distortion of the sample (hollow) on $\Delta\mathbb{T}_{\nu}/\mathbb{T}_{\nu}$ for different values of $\epsilon$.}\label{fig_hollow_1}
\vspace{1cm}
\end{figure}

\begin{figure}[ht]
\vspace{1cm}
\begin{center}
\includegraphics[width=11cm]{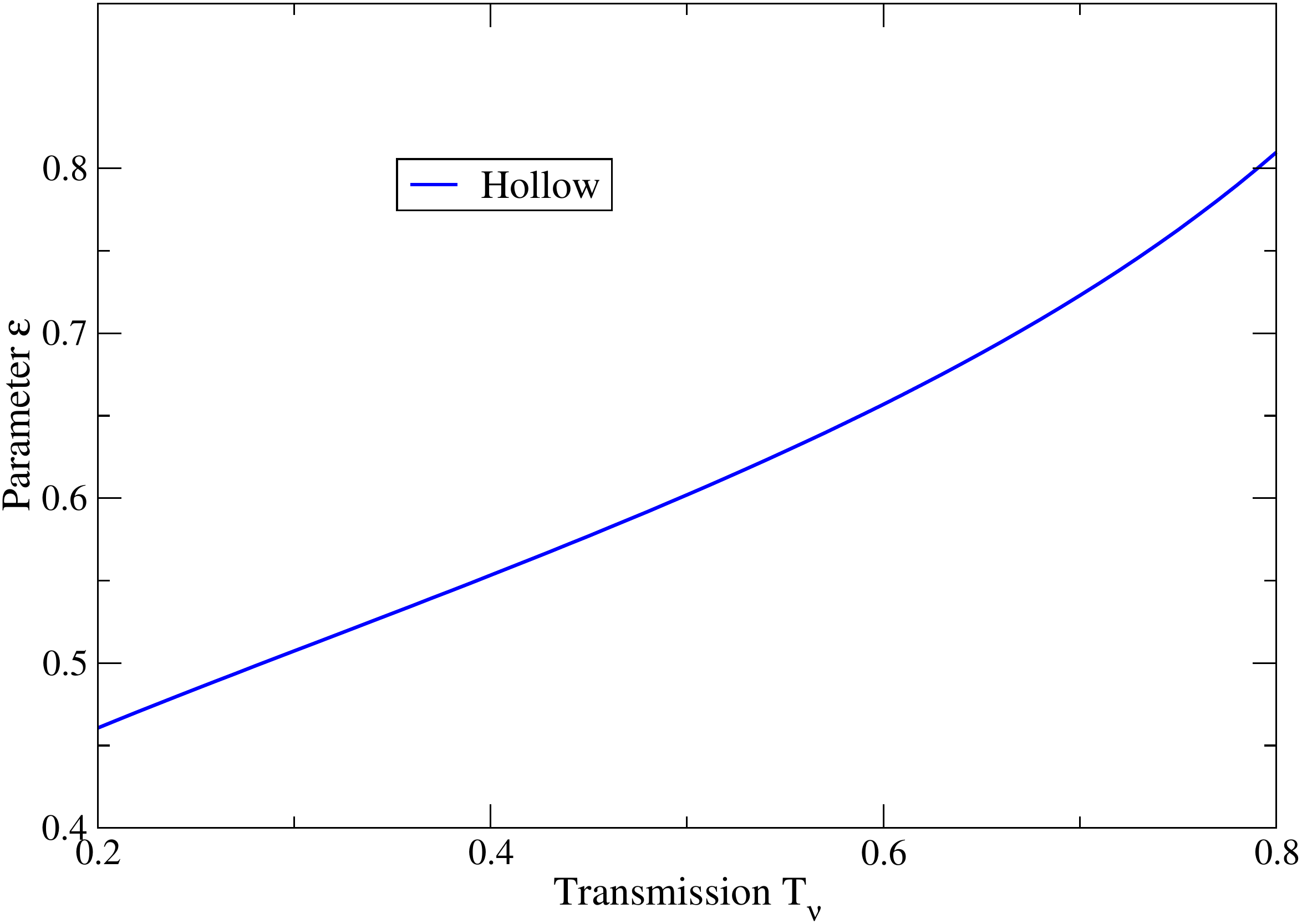}
\end{center}
\caption{(Color online) Variation of parameter $\epsilon$ as a function of transmission in the case of a hollow-shape distortion of the sample. One has $\Delta\kappa_{\nu}/\kappa_{\nu}=10 \%$ and $\Delta(\rho L)/(\rho L)=7 \%$.}\label{data3-hollow}
\vspace{1cm}
\end{figure}

\begin{figure}[ht]
\vspace{1cm}
\begin{center}
\includegraphics[width=11cm]{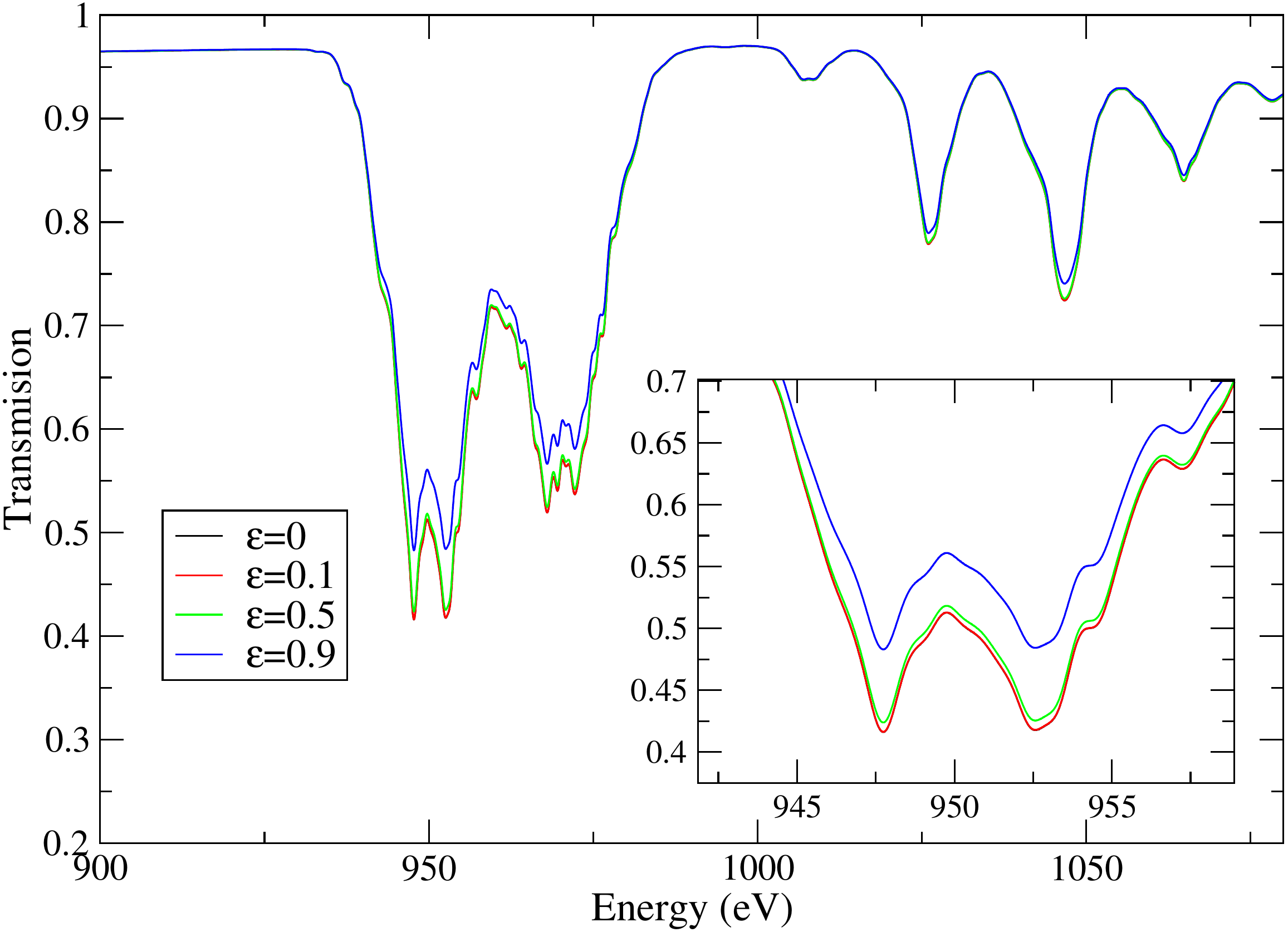}
\end{center}
\caption{(Color online) Impact of a convex distortion of the sample (hollow) on the transmission of copper at $T$=18 eV, $\rho$=0.01 g/cm$^3$ and a resolving power of $R=E/\Delta E$=1000 ($E$ is the photon energy) for different values of $\epsilon$.}\label{fig_hollow_2}
\vspace{1cm}
\end{figure}

\begin{figure*}
\vspace{1cm}
\begin{center}
\includegraphics[width=11cm]{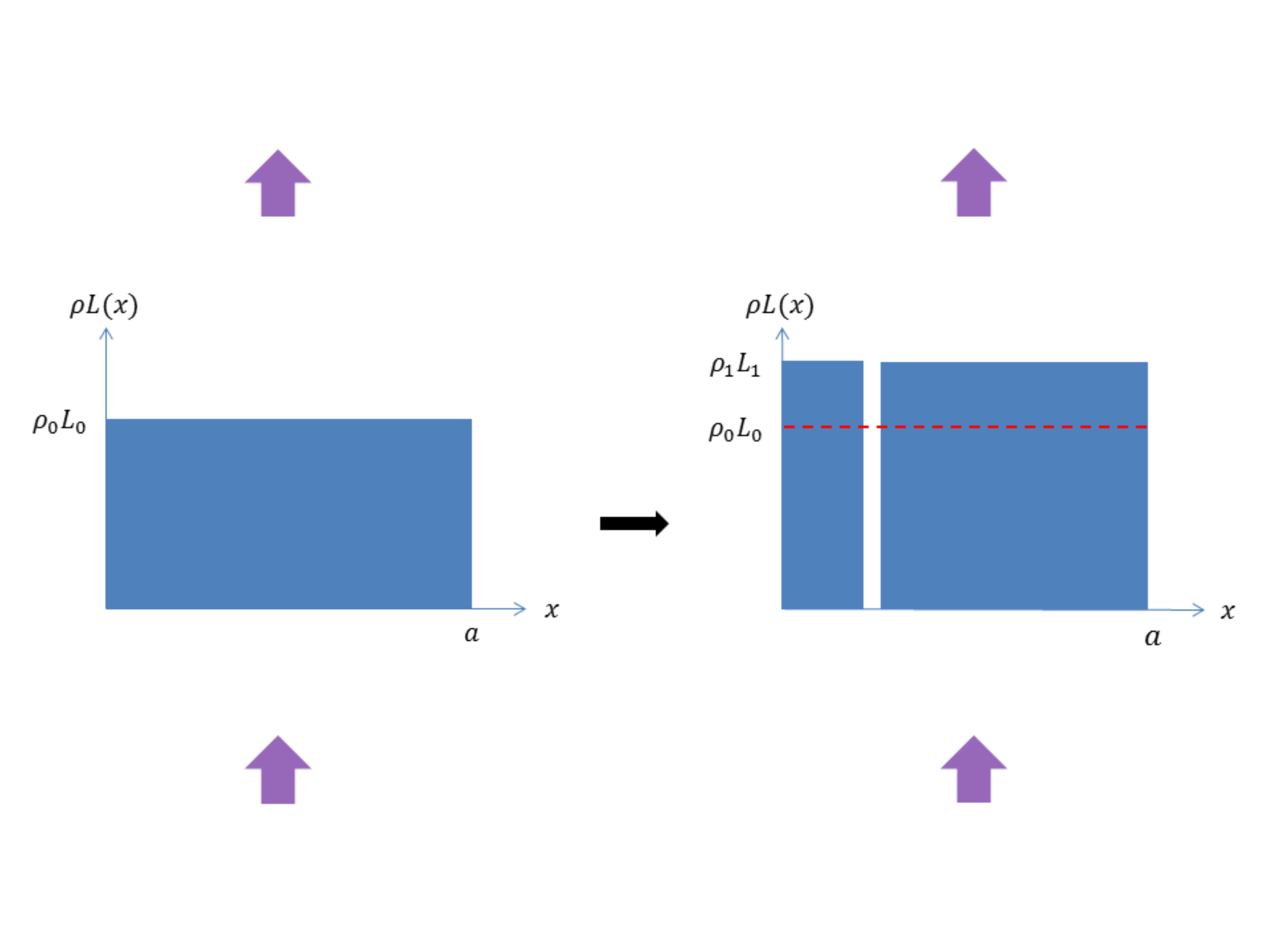}
\end{center}
\caption{(Color online) Presence of holes in the sample. The lowest purple arrows corresponds to $I_0$ and the highest to $I_{\nu}$. For the sample on the left $I_{\nu}=\mathbb{T}_{\nu}I_0$ and for the sample on the right $I_{\nu}=\tilde{\mathbb{T}_{\nu}}I_0$.}\label{holes}
\vspace{1cm}
\end{figure*}

\begin{figure}[ht]
\vspace{1cm}
\begin{center}
\includegraphics[width=11cm]{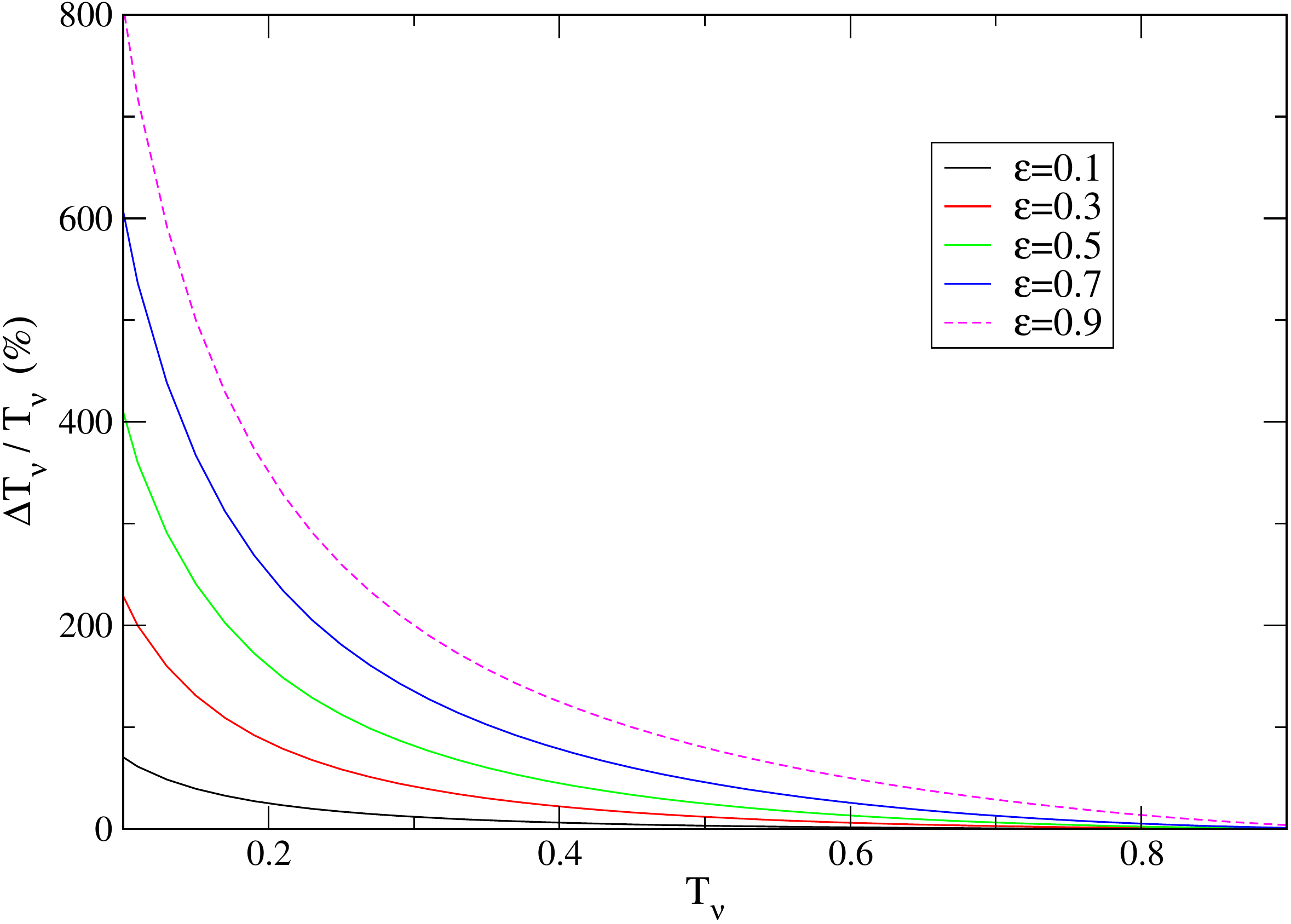}
\end{center}
\caption{(Color online) Impact of the presence of holes in the sample on $\Delta\mathbb{T}_{\nu}/\mathbb{T}_{\nu}$ for different values of $\epsilon$.}\label{fig_holes_1}
\vspace{1cm}
\end{figure}

\begin{figure}[ht]
\vspace{1cm}
\begin{center}
\includegraphics[width=11cm]{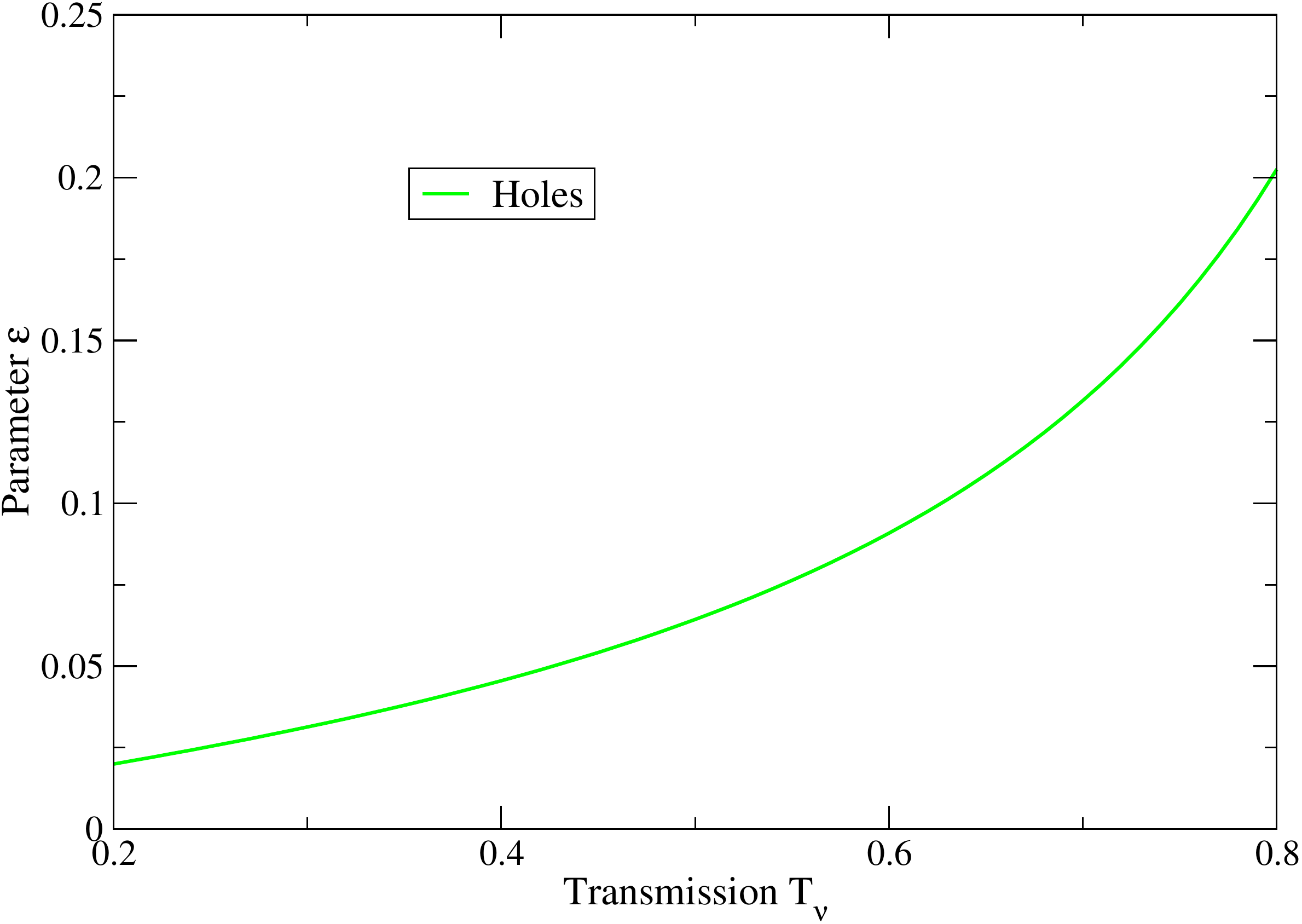}
\end{center}
\caption{(Color online) Variation of parameter $\epsilon$ as a function of transmission in the case of a sample having holes. One has $\Delta\kappa_{\nu}/\kappa_{\nu}=10 \%$ and $\Delta(\rho L)/(\rho L)=7 \%$.}\label{data4-holes}
\vspace{1cm}
\end{figure}

\begin{figure}[ht]
\vspace{1cm}
\begin{center}
\includegraphics[width=11cm]{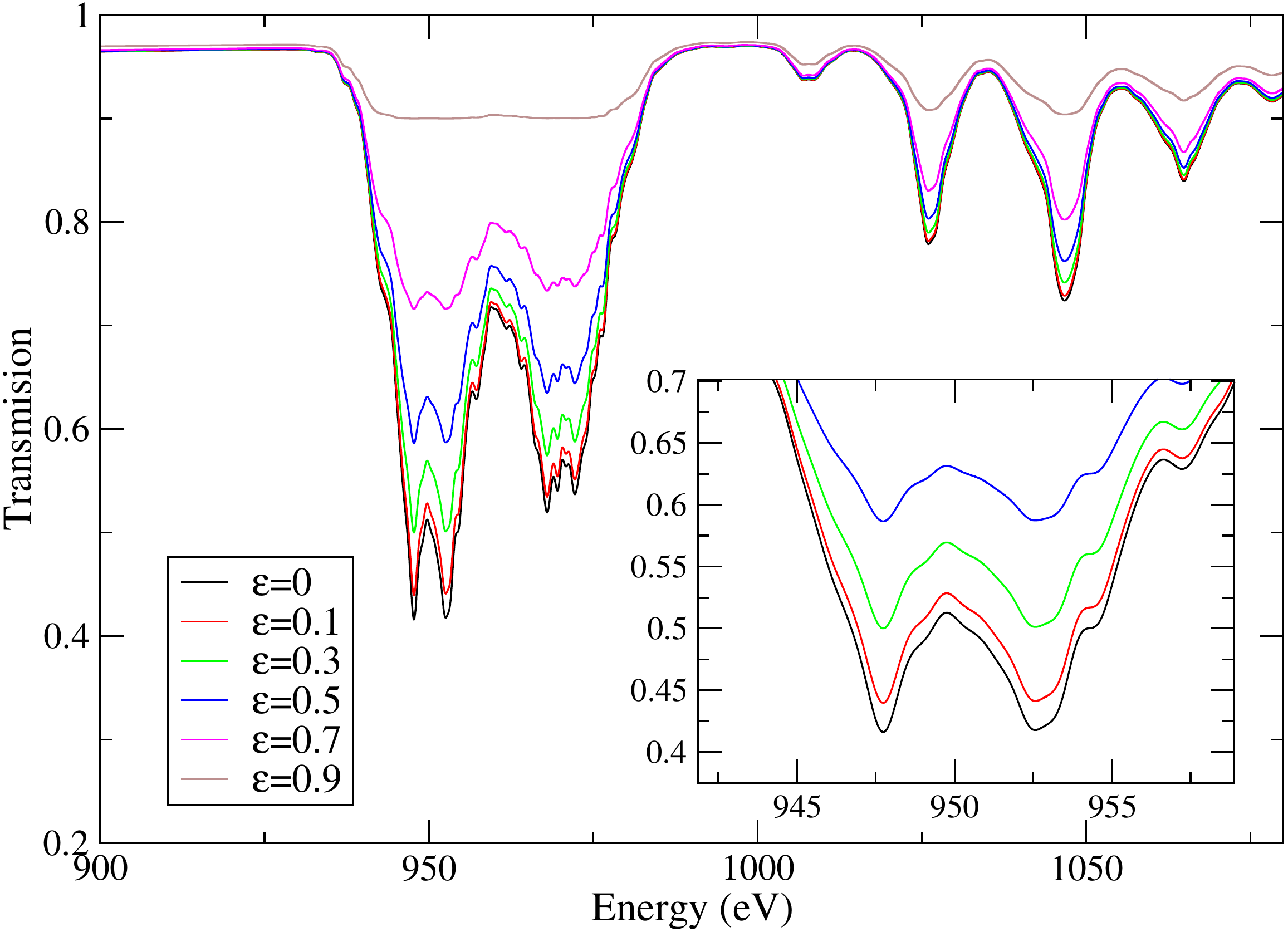}
\end{center}
\caption{(Color online) Impact of the presence of holes in the sample on the transmission of copper at $T$=18 eV, $\rho$=0.01 g/cm$^3$ and a resolving power of $R=E/\Delta E$=1000 ($E$ is the photon energy) for different values of $\epsilon$.}\label{fig_holes_2}
\vspace{1cm}
\end{figure}

\begin{figure*}
\vspace{1cm}
\begin{center}
\includegraphics[width=11cm]{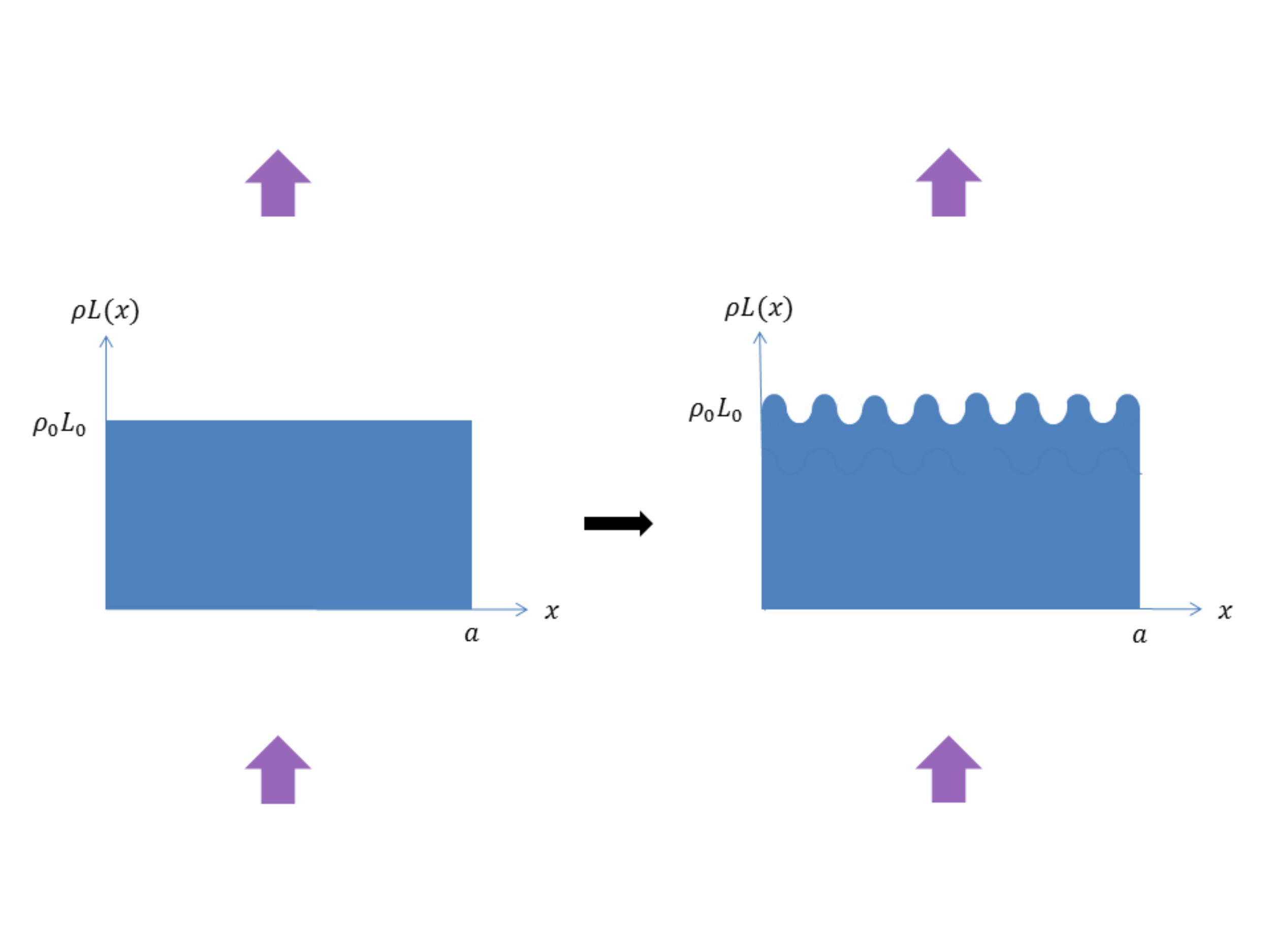}
\end{center}
\caption{(Color online) Presence of oscillations in the sample surface. The lowest purple arrows corresponds to $I_0$ and the highest to $I_{\nu}$. For the sample on the left $I_{\nu}=\mathbb{T}_{\nu}I_0$ and for the sample on the right $I_{\nu}=\tilde{\mathbb{T}_{\nu}}I_0$.}\label{oscillations}
\vspace{1cm}
\end{figure*}

\begin{figure}[ht]
\vspace{1cm}
\begin{center}
\includegraphics[width=11cm]{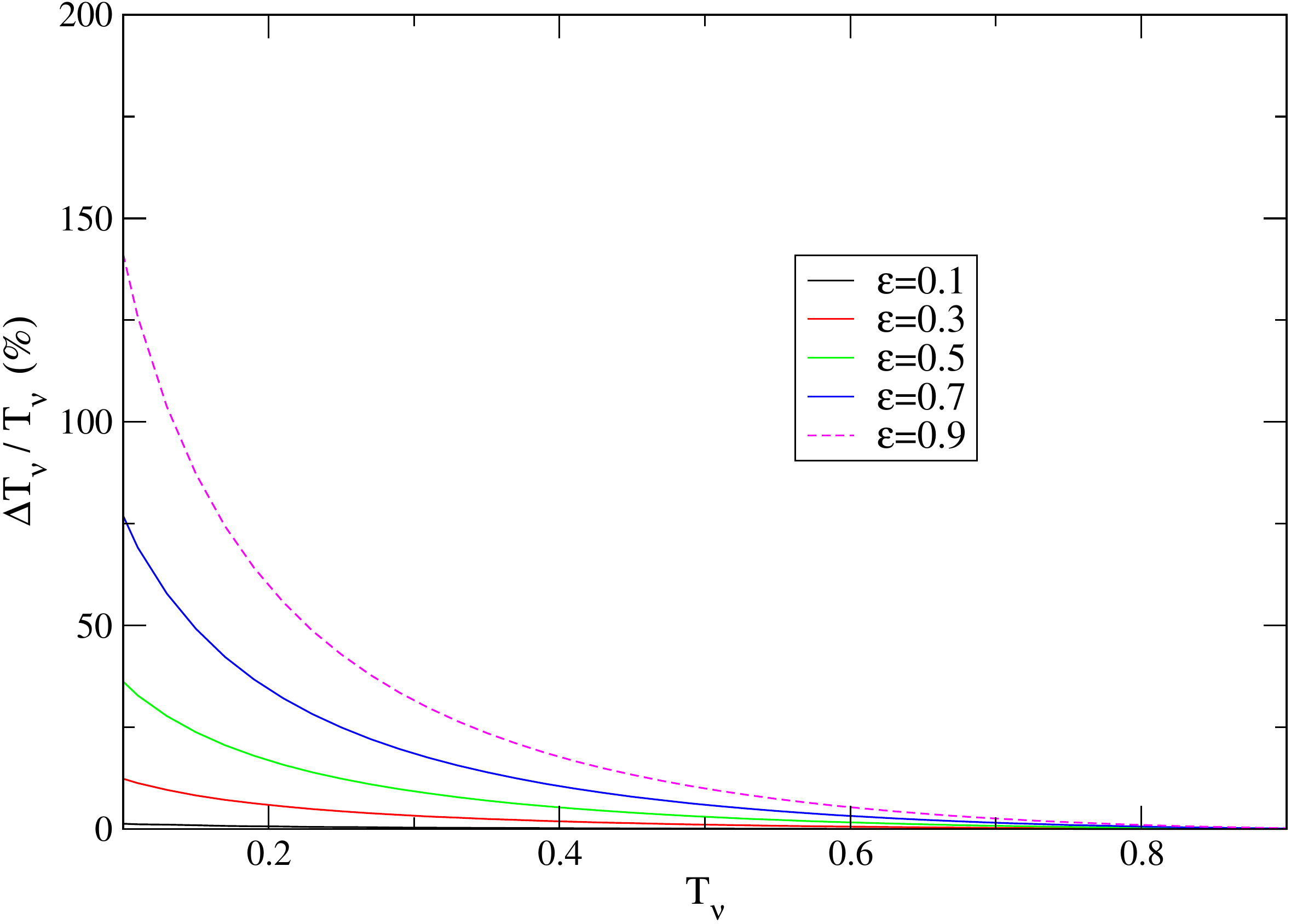}
\end{center}
\caption{(Color online) Impact of oscillations in the sample surface on $\Delta\mathbb{T}_{\nu}/\mathbb{T}_{\nu}$ for different values of $\epsilon$.}\label{fig_oscillations_1}
\vspace{1cm}
\end{figure}

\begin{figure}[ht]
\vspace{1cm}
\begin{center}
\includegraphics[width=11cm]{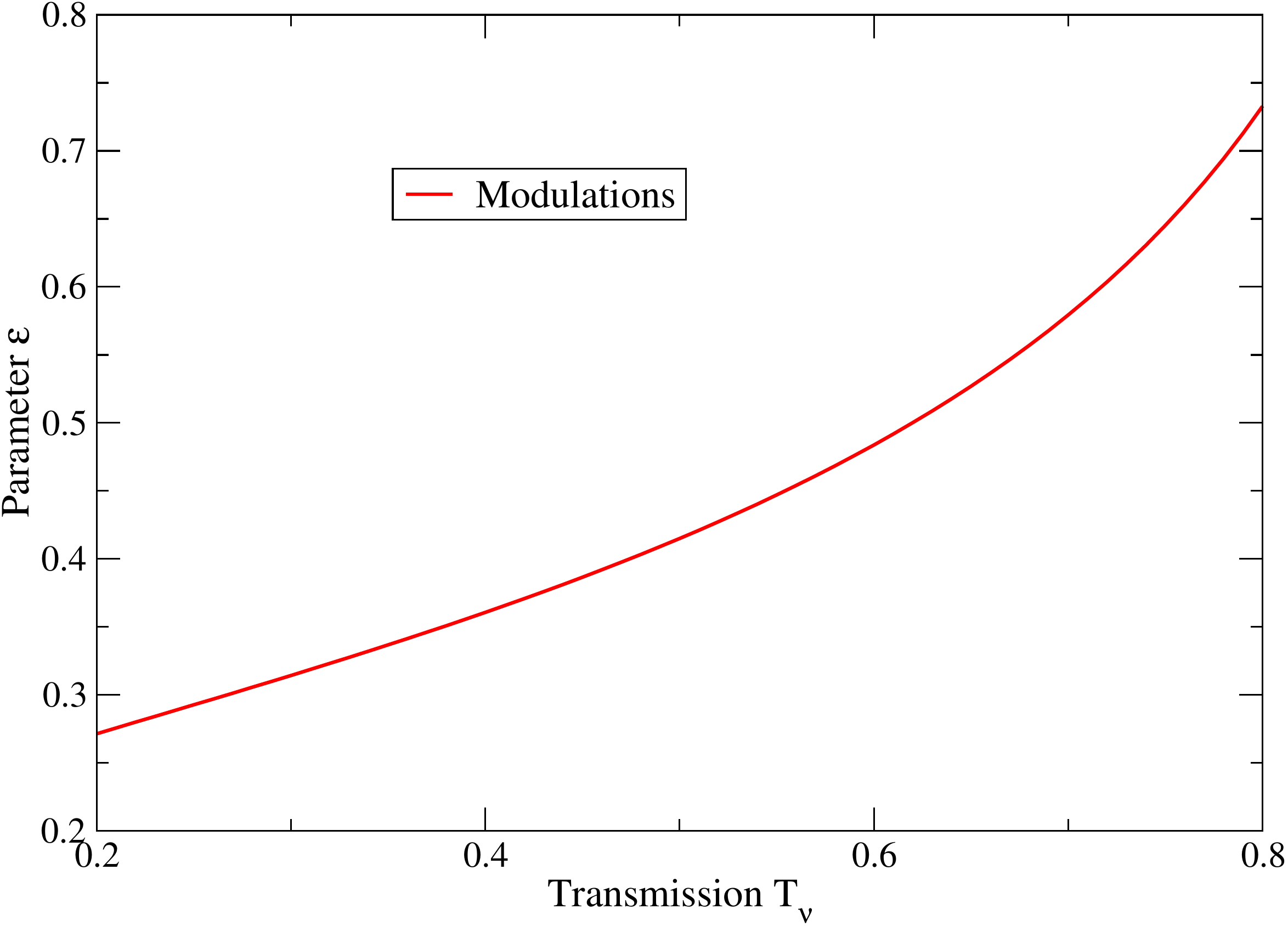}
\end{center}
\caption{(Color online) Variation of parameter $\epsilon$ as a function of transmission in the case of a sample presenting surface oscillations. One has $\Delta\kappa_{\nu}/\kappa_{\nu}=10 \%$ and $\Delta(\rho L)/(\rho L)=7 \%$.}\label{data5-modulations}
\vspace{1cm}
\end{figure}

\begin{figure}[ht]
\vspace{1cm}
\begin{center}
\includegraphics[width=11cm]{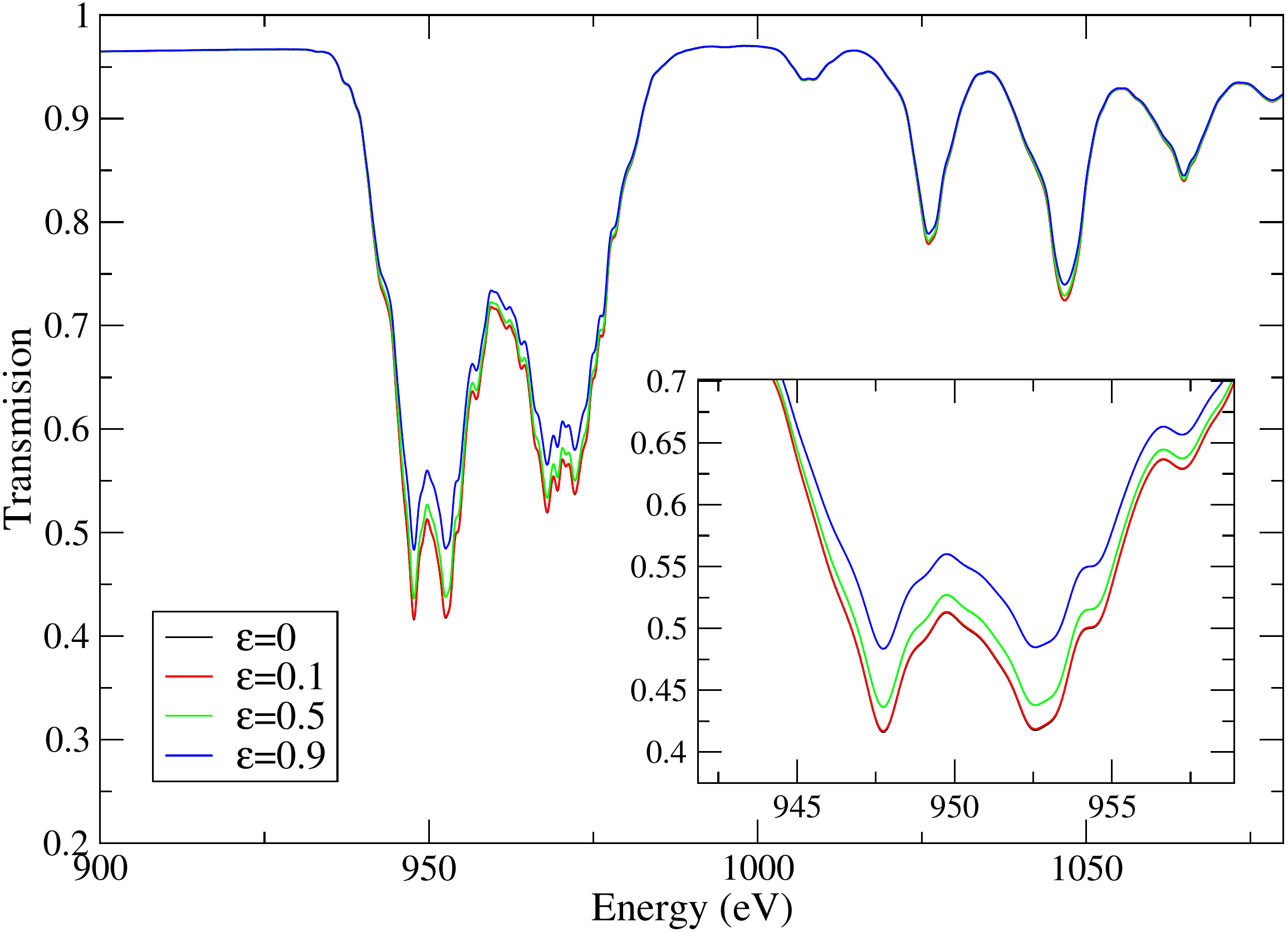}
\end{center}
\caption{(Color online) Impact of oscillations of the sample surface on the transmission of copper at $T$=18 eV, $\rho$=0.01 g/cm$^3$ and a resolving power of $R=E/\Delta E$=1000 ($E$ is the photon energy) for different values of $\epsilon$.}\label{fig_oscillations_2}
\vspace{1cm}
\end{figure}

\begin{figure*}
\begin{center}
\includegraphics[width=11cm]{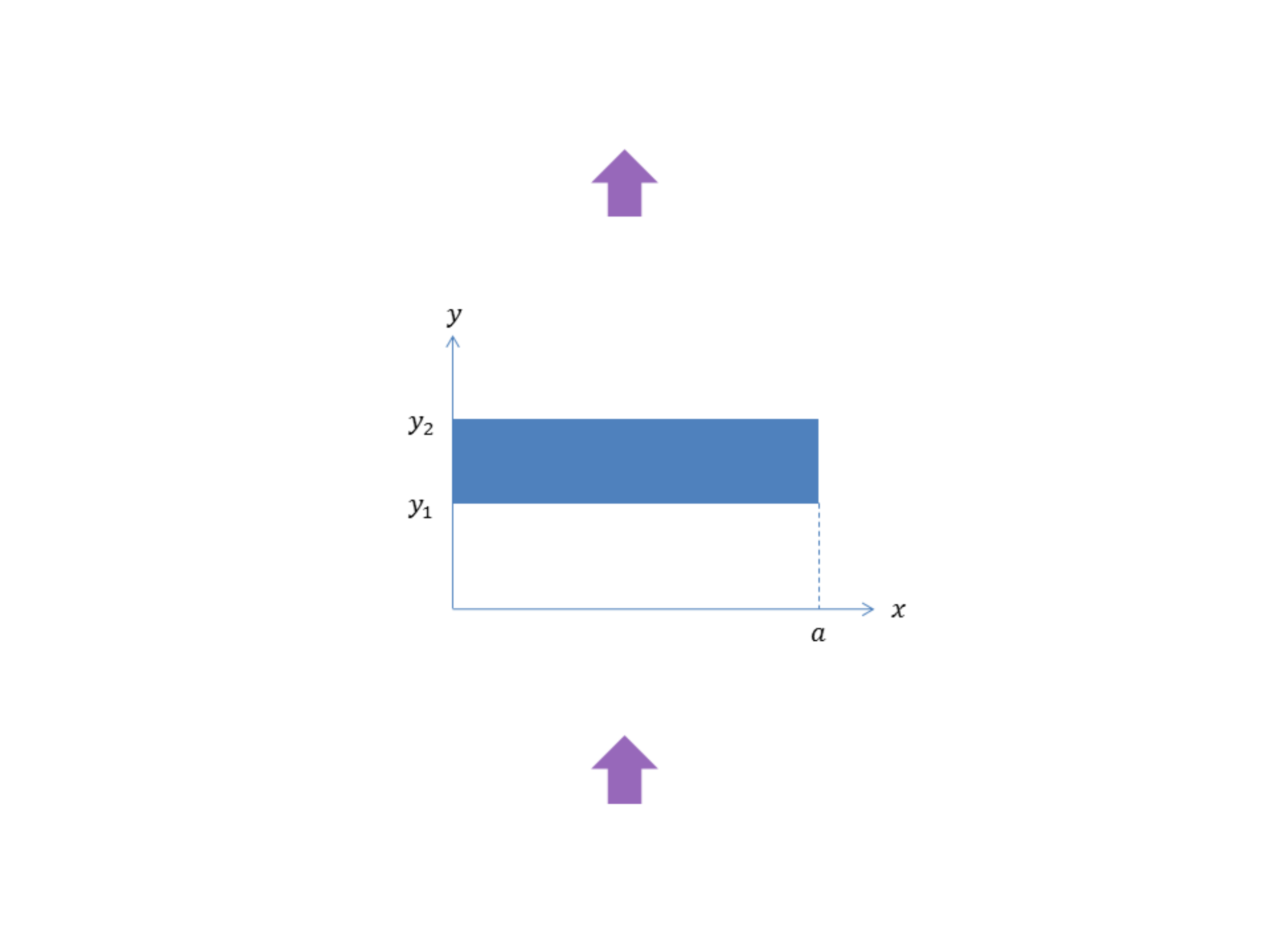}
\end{center}
\caption{(Color online) We consider a sample homogeneous along the $x$ and $z$ axis and located between $y_1$ and $y_2$ and subject to gradients along the $y$ axis.}\label{gradients}
\vspace{1cm}
\end{figure*}

\end{document}